\documentclass[aps,prl,superscriptaddress]{revtex4-1}
\usepackage{here}
\usepackage{graphicx}
\usepackage{braket}
\usepackage{color}
\usepackage{amsmath}

\begin{document}

\title{Equal-Spin Andreev Reflection in Junctions of Spin-Resolved
Quantum Hall Bulk State and Spin-Singlet Superconductor}
\author{Sadashige Matsuo}
\email{matsuo@ap.t.u-tokyo.ac.jp}
\affiliation{Department of Applied Physics, University of Tokyo, 7-3-1 Hongo, Bunkyo-ku, Tokyo
113-8656, Japan}
\author{Kento Ueda}
\affiliation{Department of Applied Physics, University of Tokyo, 7-3-1 Hongo, Bunkyo-ku, Tokyo
113-8656, Japan}
\author{Shoji Baba}
\affiliation{Department of Applied Physics, University of Tokyo, 7-3-1 Hongo, Bunkyo-ku, Tokyo
113-8656, Japan}
\author{Hiroshi Kamata}
\affiliation{Department of Applied Physics, University of Tokyo, 7-3-1 Hongo, Bunkyo-ku, Tokyo
113-8656, Japan}
\affiliation{RIKEN, Center for Emergent Matter Science, Hirosawa 2-1, Wako-shi, Saitama 351-0198, Japan}
\author{Mizuki Tateno}
\affiliation{Department of Applied Physics, University of Tokyo, 7-3-1 Hongo, Bunkyo-ku, Tokyo
113-8656, Japan}
\author{Javad Shabani}
\affiliation{California NanoSystems Institute, University of California, Santa
Barbara, CA 93106, USA}
\author{Christopher J. Palmstr\o m}
\affiliation{California NanoSystems Institute, University of California, Santa
Barbara, CA 93106, USA}
\author{Seigo Tarucha}
\affiliation{Department of Applied Physics, University of Tokyo, 7-3-1 Hongo, Bunkyo-ku, Tokyo
113-8656, Japan}
\affiliation{RIKEN, Center for Emergent Matter Science, Hirosawa 2-1, Wako-shi, Saitama 351-0198, Japan}
\begin{abstract}
The recent development of superconducting spintronics has revealed the spin-triplet superconducting proximity effect from a spin-singlet
superconductor into a spin-polarized normal metal. 
In addition recently superconducting junctions using semiconductors are in demand for highly controlled experiments to engineer topological superconductivity. 
Here we report experimental observation of Andreev reflection
in junctions of spin-resolved quantum Hall (QH) states in an InAs quantum well and the spin-singlet superconductor NbTi.
The measured conductance indicates a sub-gap feature and two peaks on the outer side of the sub-gap feature in the QH plateau-transition regime increases.
The observed structures can be explained by considering transport with Andreev reflection from two channels, one originating from equal-spin Andreev reflection intermediated by spin-flip processes and second arising from normal Andreev reflection. 
This result indicates the possibility to induce the superconducting proximity gap in the the QH bulk state, and the possibility for the development of superconducting spintronics in semiconductor devices.
\end{abstract}
\maketitle
A junction of superconductor and normal metal is a platform to observe
superconducting proximity effect, in which the superconducting
property penetrates to the normal metal. 
In a microscopic description of the proximity effect, an electron in
the normal metal enters the spin-singlet superconductor, forming a Cooper pair with another electron with opposite
spin, reflecting a hole
into the normal metal, in a process called Andreev reflection (AR)~\cite{andreevjetp1964}.
In this picture, no AR is expected in the case of a fully spin polarized normal metal, however recently, theoretical
and experimental studies in junctions with spin-polarized normal metal, revealed existence of the spin-triplet superconducting proximity effect~\cite{bergetrmp2005,keizernature2006,asanoprl2007,asanoprb2007,robinsonscience2010,visaninatphys2012,lindernatphys2015}.
The spin-triplet proximity effect is only allowed when spin-flip processes intermediate "equal-spin" AR, which is possible due to the presence of magnetization inhomogeneity or spin-orbit interaction~\cite{eschrignatphys2008,ostaayprb2011}.
Ferromagnetic metal has been utilized in experiments of superconducting
spintronics to date. 
A semiconductor material however offers several advantages, including the control of carrier density and spin filling through electrical gating and magnetic field, and the possibility of ballistic transport in micron sized devices. 
Furthermore strong spin-orbit interaction can be utilized in two dimensional electron gases (2DEGs) in narrow gap semiconductors such as InAs and InSb~\cite{luoprb1988,silvaprb1997,nittaprl1997,heidaprb1998}.
These features favor the formation of spin-polarized states when the 2DEG is in the quantum Hall (QH) regime.
Indeed, there are several experimental reports focusing on superconductor-semiconductor
junctions in the QH regime~\cite{takayanagiphysb1998,eromsprl2005,rickhausnl2012,komatsuprb2012,wannatcommun2015,ametscience2016}.
However, all of these experiments have focused on the
spin-degenerate QH states
and the spin-triplet proximity effect has yet to be experimentally addressed, despite theoretical predictions of spin-triplet supercurrent in Josephson junctions with weak links of spin resolved QH edge channels~\cite{ostaayprb2011}. 
Additionally, if the superconducting proximity gap is induced into the spin-resolved QH state, the system can be a topological superconductor~\cite{xiaoprb2010} and give a new platform to realize the Majorana Fermions~\cite{fuprl2008,hasanrmp2010} whose signatures have recently been reported~\cite{mourikscience2012, dasnatphys2012,  rokhinsonnatphys2012, nadjpergescience2014, boucquillonnatnano2016, wiedenmannnatcommun2016, albrechtnat2016, deaconarxiv2016}.

Here we report an experimental study on electron transport in junctions
between spin-resolved QH states and spin-singlet superconductors. We
prepared junctions consisting of a high mobility InAs quantum well (QW) with NbTi contacts. 
The NbTi layers are contacted to the sides of the mesa containing the QW, minimizing the damage to the 2DEG.
The 2DEG possesses a large g-factor, high mobility, and strong spin-orbit interaction, all necessary ingredients for coexistence of superconductivity and spin-resolved QH states. 
We observe spin-resolved quantized steps at magnetic fields below the superconducting critical field, 7 T, and find that the differential conductance has a dip or a peak structure as a sub-gap feature in all QH plateau-transition regimes of filling factor between 0 to 4. 
Additionally, we find two side peaks on the outer side of the sub-gap feature. 
We conclude that the structures observed here are a result of the equal-spin AR between the spin-resolved QH bulk states
and the superconductor.

\begin{figure}[t]
\centering 
\includegraphics[width=.6\linewidth]{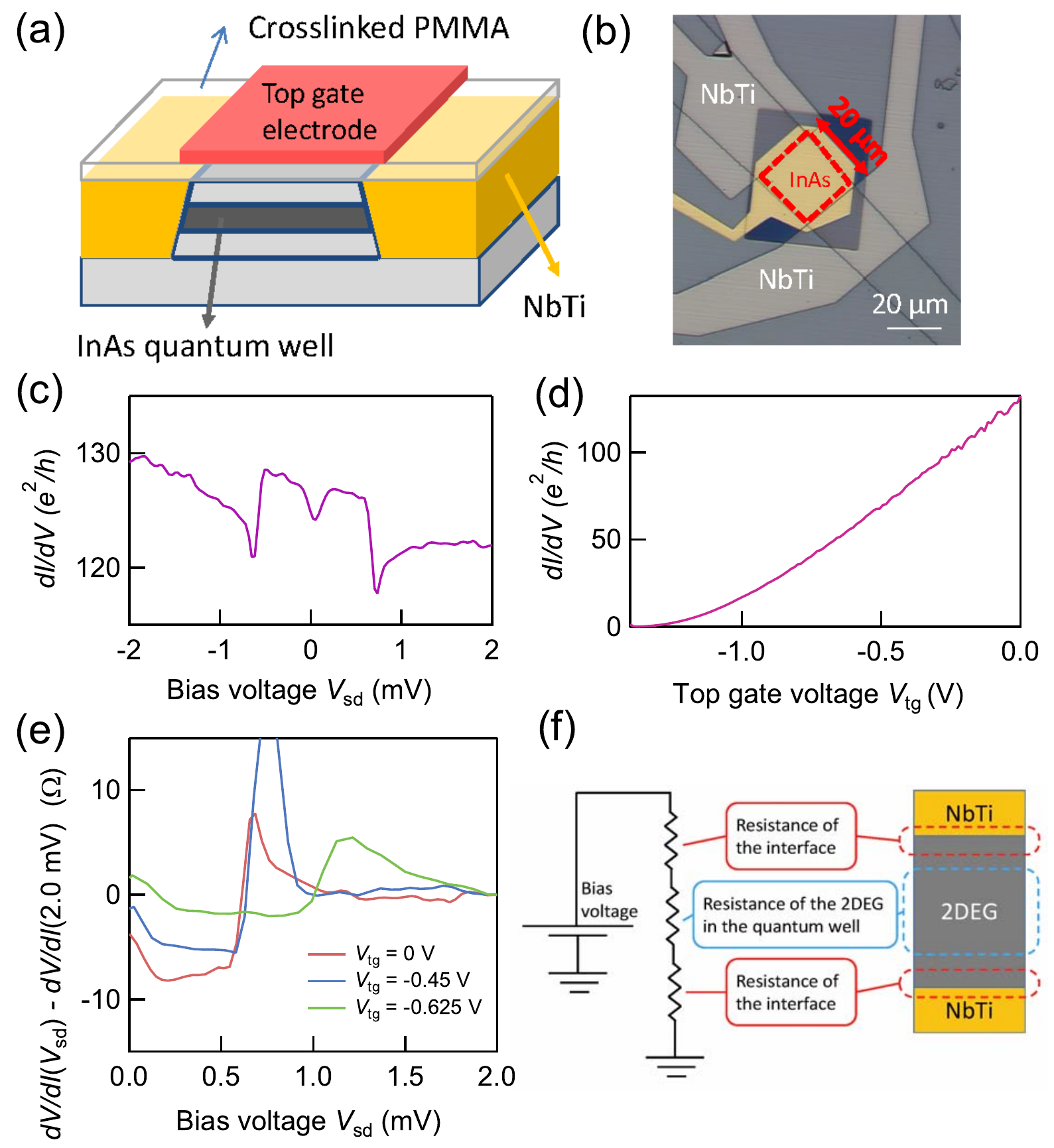}
\caption{(a)Cross section of the fabricated device. The edges of the InAs QW are contacted with sputtered NbTi. (b)Optical image of the device.
The region surrounded by the red dash line represents the mesa. 
(c)$dI/dV$ vs. $V_{{\rm sd}}$ at $V_{{\rm tg}}=0$ V and $B=0$~T.
$dI/dV$ measured in the range $-0.71~{\rm mV}<V_{{\rm sd}}<0.71~{\rm mV}$
is enhanced due to AR. 
(d)$dI/dV$ vs. $V_{{\rm tg}}$ at $V_{{\rm sd}}=0$ V
is shown. The InAs QW is completely depleted by $V_{{\rm tg}}$. 
(e)$dV/dI$ with $dV/dI$ measured at $V_{{\rm sd}}=2.0$~mV subtracted
as a function of $V_{{\rm sd}}$ is shown. The red, blue, and green
lines are measured at $V_{{\rm tg}}=$0, -0.45, and -0.625~V, respectively.
The resistance reduction due to AR decreases as $V_{{\rm tg}}$ decreases, indicating that $V_{{\rm tg}}$ tunes the carrier density of not only the center region of the
2DEG but also region near the junctions. (f)Schematic image of an
equivalent circuit to our junction devices. The applied $V_{{\rm sd}}$
is divided between the two junctions and the 2DEG.
}
\label{fig1} 
\end{figure}

We fabricated junction devices from an InAs QW grown by molecular beam epitaxy~\cite{shabaniprb2014,shabaniapl2014} with carrier density $3\times10^{11}~{\rm cm^{-2}}$
and mobility $3\times10^{5}~{\rm cm^{2}V^{-1}s^{-1}}$. (The material stack is represented in Supplemental Material (SM).) 
The cross section of the device is schematically represented
in Fig.~\ref{fig1}(a). 
Sputtered NbTi with the critical field of 7.0 T and the critical temperature of 6.5 K contacts the edge of the 2DEG and a top gate structure is fabricated using an insulating layer of cross-linked PMMA (see SM for details).
The optical microscope photo in Fig.~\ref{fig1}(b) shows the top view of the device. 
The two junctions are separated by 20 ${\rm \mu}$m, so this device is assumed to have two independent junctions. 
We measured the two-terminal differential
conductance in a He3He4 dilution refridgerator with a base temperature of 50 mK.

Figure~\ref{fig1}(c) shows the measured $dI/dV$ vs. $V_{{\rm sd}}$ at $B=$0 T.
The $dI/dV$ is enhanced in the range of $-0.71~{\rm mV}<V_{{\rm sd}}<0.71$~mV.
This conductance enhancement arises from AR~\cite{andreevjetp1964, BTK}
and only appears for the bias voltage in the junction less than the
superconducting bulk gap energy $\Delta_{bulk}$. Therefore, we
evaluate the $\Delta_{bulk}$ as approximately 0.35 meV. The observation
of the sub-gap feature guarantees that the junction has enough quality
to study AR between the superconductor
and spin-resolved QH state.

Figure~\ref{fig1}(d) shows $dI/dV$ vs. the top
gate voltage, $V_{{\rm tg}}$. $dI/dV$ gradually decreases with
decreasing $V_{{\rm tg}}$ and becomes pinched off for $V_{{\rm tg}}<$~-1.38
V. This indicates that the top gating efficiently varies the 2DEG
carrier density but not necessarily near the junction. We measured $dI/dV$ vs. $V_{{\rm sd}}$ for various
values of $V_{{\rm tg}}$ to examine the effect of top gating on the sub-gap conductance. We derived the differential resistance
as given by $dV/dI=(dI/dV)^{-1}$ and then subtracted $dV/dI$ measured at $V_{{\rm sd}}=2.0$~mV to
eliminate the normal state resistance including a series resistance
due to the InAs QW away from the junction.
The result is shown in Fig.~\ref{fig1}(e) for three different values
of $V_{{\rm tg}}$ =0, -0.45, and -0.625 V by the purple, blue, and
green curve, respectively. It is clear to see that $dV/dI(V_{{\rm sd}})-dV/dI(2.0~{\rm mV})$
displays a pronounced reduction (or conductance enhancement) within the gap, and the reduction increases as $V_{{\rm tg}}$ is made more positive.
If the top gating only varies the carrier density away from the junction,
the $dV/dI$ reduction below the gap should be constant with $V_{{\rm tg}}$.
Therefore, the result of Fig.~\ref{fig1}(e) indicates that the top gating
is efficient enough to vary the carrier density in the InAs QW near the superconducting junction. In Fig.~\ref{fig1}(e), $V_{{\rm sd}}$
to characterize the superconducting gap decreases as $V_{{\rm tg}}$ is made more positive.
The $V_{{\rm sd}}$ shift is due to the change in the voltage dropped over the junctions when the series resistance of the 2DEG is altered. Schematics of the
equivalent circuit for the device are shown in Fig.~\ref{fig1}(f).
In the constant voltage bias measurement, the effective junction voltage
decreases as the carrier density of the QW decreases with decreasing $V_{{\rm tg}}$. Herein the applied voltage of $V_{{\rm sd}}$ is assumed to only drop
across the junctions in the saturation region. Therefore we evaluate $\Delta_{bulk}\simeq0.35$
meV. This $\Delta_{bulk}$ is smaller than the value of 0.99 meV, predicted from the critical temperature with conventional BCS theory.
A reduced $\Delta_{bulk}$ has been reported in previous studies~\cite{irieprb2016}, and assigned to the normal reflection in the junction as a possible origin. 
The normal reflection can lift the degeneracy of the gap, making the energy lower~\cite{tangzpb1996}. 

\begin{figure}[t]
\centering 
\includegraphics[width=.6\linewidth]{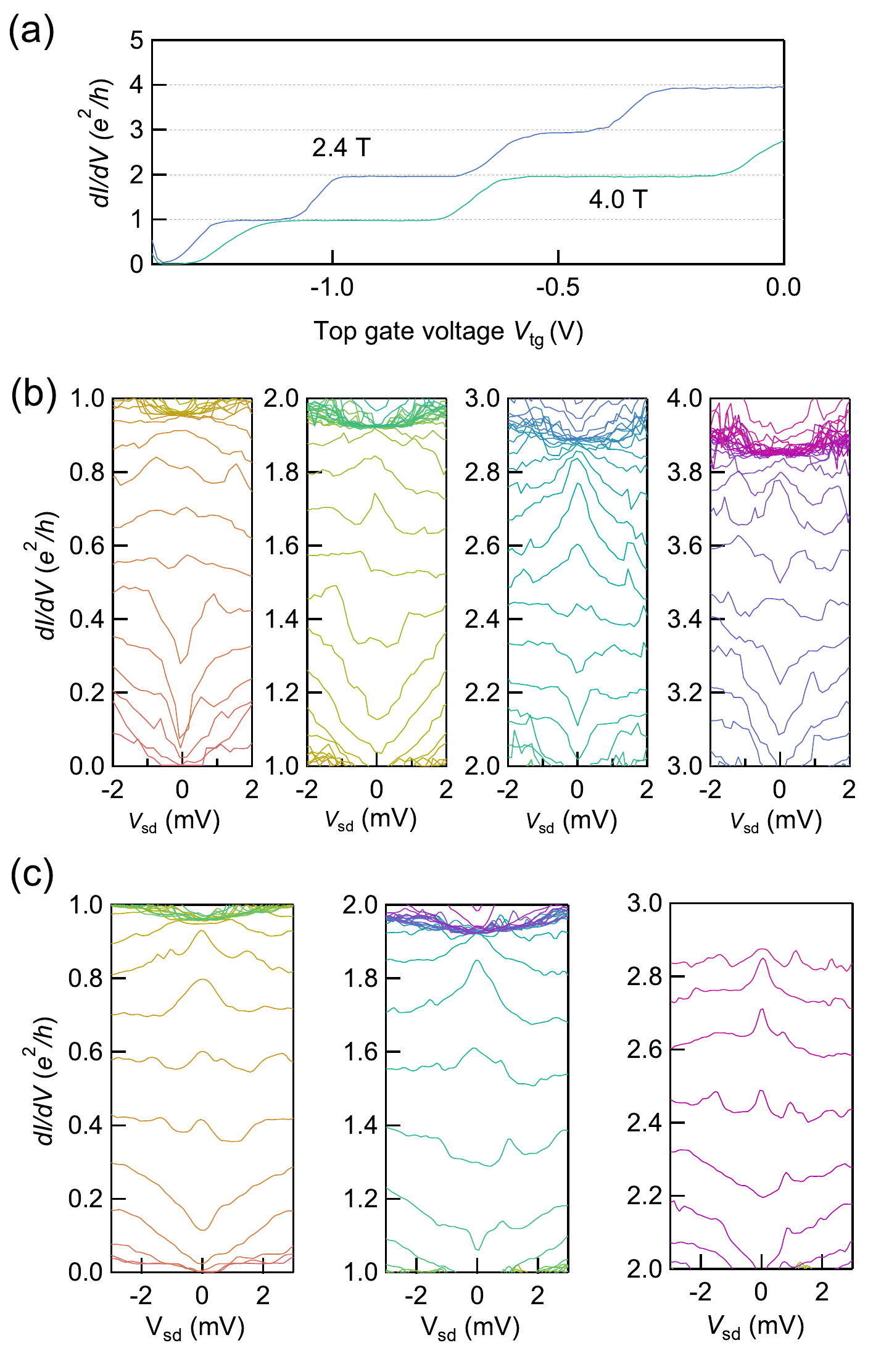} 
\caption{
(a)$dI/dV$ vs. $V_{{\rm tg}}$ at 2.4 T and 4.0 T with $V_{{\rm sd}}=0$~V. The
conductance plateaus are clearly observed on 1, 2, 3, and 4$\times e^{2}/h$.
This indicates the Zeeman energy at 2.4 T is enough to resolve the
spin degeneracy. (b)Measured $dI/dV$ vs. $V_{{\rm sd}}$ at 2.4~T
for -1.5~V$<V_{{\rm tg}}<0$~V, divided into four panels to clarify
the $V_{{\rm sd}}$ dependence in each plateau-transition regime. $dI/dV$
has a dip structure around $V_{{\rm sd}}=0$ V for $ne^{2}/h<dI/dV<ne^{2}/h+0.5$.
In contrast $dI/dV$ has a peak structure for $ne^{2}/h+0.5<dI/dV<ne^{2}/h+0.8$.
(c)Measured $dI/dV$ vs. $V_{{\rm sd}}$ at 4.0~T for -1.5~V$<V_{{\rm tg}}<0$~V divided into three panels. The features are similar to those observed in the results obtained at 2.4 T. }
\label{fig2} 
\end{figure}

In Fig.~\ref{fig2}(a), we present plots of measured $dI/dV$ as a function of $V_{{\rm tg}}$ at out-of-plane
magnetic field $B=2.4$ and $4.0$ T. The well-defined
plateaus at integer multiples of $e^{2}/h$ which originate from the QH
edge transport are clearly seen. From this observation, it is confirmed that the applied
fields are strong enough to resolve the spin degeneracy, but significantly smaller than the critical field of the NbTi,
implying that the superconductivity and the spin-resolved QH states
coexist. 

$dI/dV$ vs. $V_{\rm sd}$ measured in the range of $V_{{\rm tg}}$ between -1.5 V
and 0 V is represented in Fig.~\ref{fig2}(b), and (c) for $B=2.4$
T, and $4.0$ T, respectively. $dI/dV$ traces measured for magnetic fields spanning the conductance range $ne^{2}/h$ to $(n+1)e^{2}/h$ with $n=0,1,2...$ are shown in the
separate panels to highlight the sub-gap structure in the transition regions between plateaus. 
For example, $dI/dV$ at $B=2.4$ T between $0<dI/dV<e^{2}/h$ is shown in the leftmost panel of Fig. 3(b). 

In the transition regions between the conductance plateaus, we find pronounced sub-gap
features appearing as a dip, a peak and then a dip at around $V_{\rm sd}$
$=0$ V from the lower to the upper plateau in all panels. Similar
sub-gap features are previously reported for the junctions of two
NbN superconductors and a spin-degenerated QH state in an InAs QW~\cite{takayanagiphysb1998}. However, the underlying physics of the sub-gap feature
remains to be elucidated. More interestingly in Fig.~\ref{fig2}(b) and (c),
the center peak appears broad in some traces and even split into two
in others. In addition to a center dip or peak structure we observe a side peak. For
example, it is clear to see two side peaks at $V_{\rm sd}=\pm 1.8$ mV
in addition to a peak at $V_{\rm sd}=0$ V for the curves at $dI/dV \simeq 2.5e^{2}/h$
in the right panel of Fig.~\ref{fig2}(c).

\begin{figure}[t]
\centering 
\includegraphics[width=.6\linewidth]{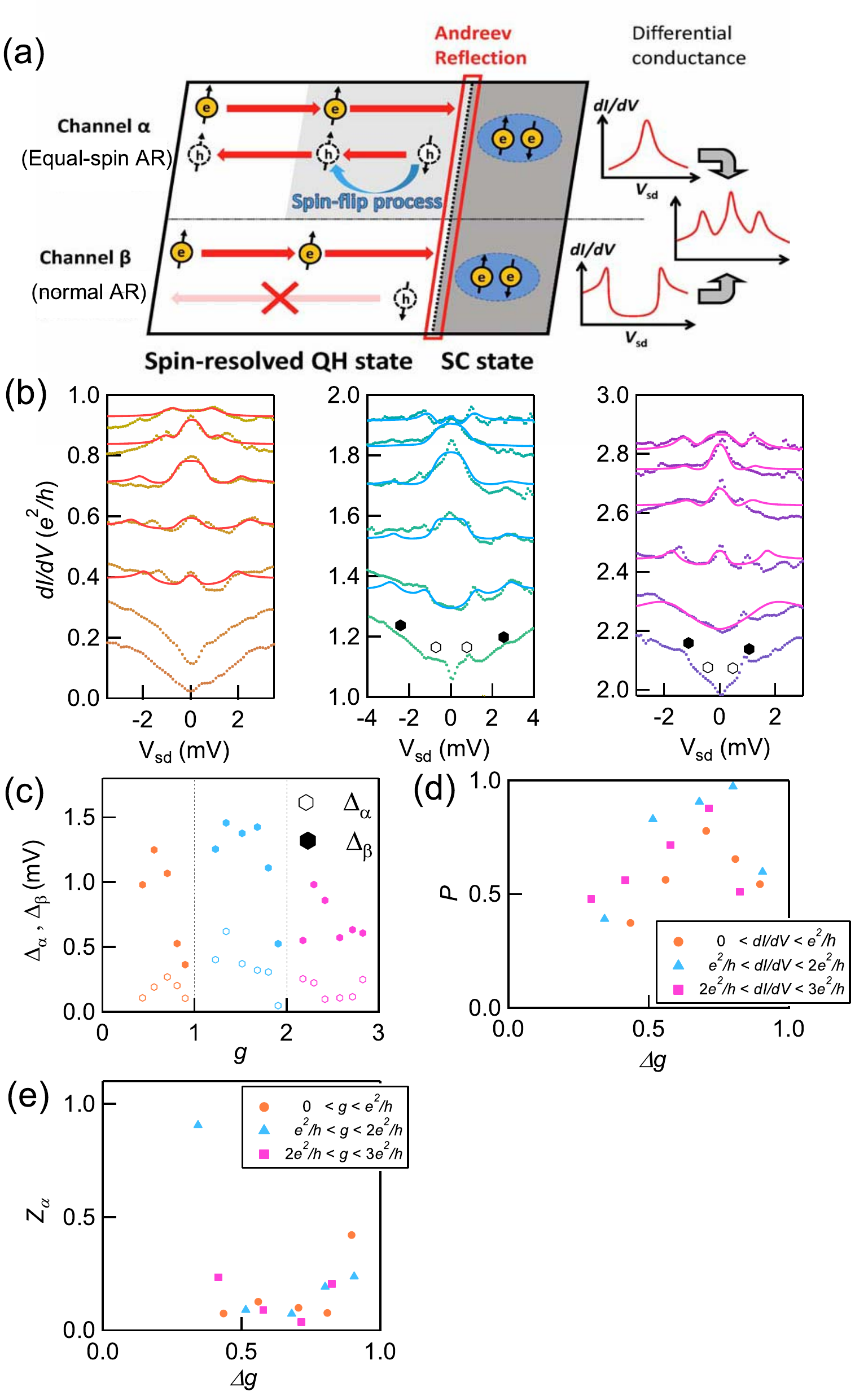}
\caption{(a)Schematic of AR in the channel $\alpha$ and $\beta$. AR in channel $\alpha$ is intermediated by the spin-flip process, while the reflection in the channel $\beta$ is not. (b)Measured $dI/dV$ vs. $V_{{\rm sd}}$ at 4.0 T indicated with dots with fitting results shown as solid lines. The open and closed hexagons are the position of $\Delta_{\alpha}$ and $\Delta_{\beta}$ without the numerical fitting. (c)Obtained $\Delta_{\alpha}$ and $\Delta_{\beta}$ are shown as open and closed hexagons, respectively. $\Delta_{\beta}$ has a convex upward trend in the respective plateau-transition regime. (d)$P$, indicating the relative contribution of the two channels, is shown as a function of $\Delta g$. The orange circles, blue triangles and purple squares are $P$ calculated from the analysis of the left, center, and right panels. (e)$Z_{\alpha}$ is shown as a function of $\Delta g$. The position where $Z_{\alpha}$ has the minimum is the same as the $P$. }
\label{fig3} 
\end{figure}

The sub-gap conductance enhancement indicated by the observation of a zero-bias
peak can be assigned to AR in the junction having a low potential barrier. In contrast, if the
potential barrier is so large that normal reflection is more dominant
than AR, a dip rather than a peak can appear according
to the Blonder, Tinkham, and Klapwijk (BTK) theory~\cite{BTK}. Therefore
we assign the peak (dip) structure observed in the plateau-transition
regime to AR (normal reflection), and then the
change of the sub-gap feature in Fig.~\ref{fig2} (b) and (c) can
be simply explained by considering the change of the junction potential
barrier depending on $V_{\rm tg}$. For the transport through
the QH state it is well established that the dominant contribution
arises from the QH bulk state in the plateau-transition regime and
from the QH edge state in the plateau regime. Herein, we deduce that
the sub-gap feature, especially the peak structure is originated from AR in the junction between the superconductor
and the spin-resolved QH bulk state. A finite amount of equal-spin AR can be expected for a 2DEG with strong spin-orbit interaction according to recent theoretical
studies~\cite{eschrignatphys2008,ostaayprb2011}. Therefore, we here assume coexistence of equal-spin AR and normal AR, corresponding to AR intermediated with and without a spin-flip process respectively, as schematically shown in Fig.~\ref{fig3}(a).

In order to more quantitatively interpret the sub-gap features including
the side peaks, we construct a model based on the BTK theory. In this
theory, the normalized differential conductance of the junction, $G_{\rm int}(V_{\rm sd},Z,T,\Delta)$,
can be written as
\begin{eqnarray}
G_{\rm int}(V_{\rm sd},Z,T,\Delta)=\int_{-\infty}^{\infty}\frac{df(E-V_{\rm sd},T)}{dV_{\rm sd}}(1+A(E,Z,\Delta)-B(E,Z,\Delta))dE &  &, \label{BTK}
\end{eqnarray}
where $f(E),T,\Delta$ and $Z$ are the Fermi-Dirac distribution function,
temperature, superconducting gap and the dimensionless parameter representing
the potential barrier in the junction, respectively. $A$($B$) is
the probability of the equal-spin AR (normal reflection) defined
in the BTK theory~\cite{BTK}. However, this standard BTK theory
cannot explain the coexistence of side peaks and sub-gap feature as observed in Fig.~\ref{fig2}(b)
and (c). Herein, to apply the BTK theory for such cases,
we assume two different transport channels in the proximity region,
labeled channel $\alpha$ in which the equal-spin AR occurs and
channel $\beta$ with no spin-flip process. A schematic representation of the transport in these channels is shown in Fig.~\ref{fig3}(a). The
channel $\alpha$ can generate the sub-gap features reflecting the
conductance enhancement due to the equal-spin AR, while the channel $\beta$
can generate the side peaks reflecting the quasiparticle peaks of the superconducting bulk gap energy via normal reflection. Then, the normalized differential conductance of the junction
can be written as 
\begin{eqnarray}
P\times G_{{\rm int}}^{\alpha}(V_{{\rm sd}},Z_{\alpha},T,\Delta_{\alpha})+(1-P)\times G_{{\rm int}}^{\beta}(V_{{\rm sd}},Z_{\beta},T,\Delta_{\beta}),
\end{eqnarray}
where $G_{{\rm int}}^{\alpha},G_{{\rm int}}^{\beta},Z_{\alpha},{\rm and}Z_{\beta}$
are the normalized differential conductance of the channel $\alpha$,
the normalized differential conductance of the channel $\beta$, the
parameter $Z$ in the eqn.~(\ref{BTK}) of the channel $\alpha$ and
that of the channel $\beta$, respectively. $\Delta_{\alpha}$, and $\Delta_{\beta}$
are the proximity gap energy, and the bulk gap energy, respectively. Parameter $P$ indicates the relative contribution of the two AR channels.
We note that the appearance of two superconducting gaps are theoretically discussed in the case of coexistence of spin-singlet and triplet superconducting pair amplitude~\cite{bursetprb2014, bursetprb2015}.
We executed numerical calculations to fit the experimental data (see SM).

The best fitting result is shown in Fig.~\ref{fig3}(b) by the solid
lines plotted alongside the experimental data at 4.0 T. The obtained $\Delta_{\alpha}$
and $\Delta_{\beta}$ are plotted as a function of
$dI/dV$ in Fig.~\ref{fig3}(c). $dI/dV$ for the x-axis is the
differential conductance of the normal state measured at $V_{\rm sd}=3.5$~mV in units of $e^{2}/h$ and indexed by $g$. 
$\Delta_{\beta}$ is derived from the side peak positions and has
a convex upward trend in each plateau-transition regime between plateaus
of g= 0 and 1, 1 and 2, and 2 and 3, respectively. These $\Delta_{\beta}$ values
are larger than the true bulk gap due to the dissipation in the
QH bulk state (the equivalent circuit is the same as shown in Fig.~\ref{fig1}(f)).
We assume that the superconducting bulk gap $\Delta_{bulk}=0.35$
meV which is derived in Fig.~\ref{fig1}(c) is unchanged with $V_{\rm tg}$
and therefore $\Delta_{\beta}$ in Fig.~\ref{fig3}(c) should
be equal to $\Delta_{bulk}$. This assumption is probably valid because the observed $\Delta_{\beta}$ in the plateau regime where there is no dissipation
is consistent with $\Delta_{bulk}$ (see SM).
Then we use the same equivalent circuit model as used for evaluating $\Delta_{bulk}$
to calibrate the value of $\Delta_{\rm \alpha}$ and finally obtain
the true proximity gap of $\Delta_{\rm triplet}\simeq 0.1$
meV as $0.35\times\Delta_{\alpha}/\Delta_{\beta}$ shown in Fig.~\ref{fig3}(c)
(see SM for details). 

We also derived the parameters $P$ and $Z_{\alpha}$ and plot them
as a function of change of $g$ in Fig.~\ref{fig3}(c), i.e. $\Delta g=0$ to $1$
between plateaus in Fig.~\ref{fig3}(d) and (e), respectively. The
pink rectangles, blue triangles, and orange circles represent the
parameters derived from the right, center, and left panel of Fig.~\ref{fig3}(a),
respectively. $P$ indicating the proportion of AR in channel $\alpha$ has the
maximum ($\simeq1$) and $Z_{\alpha}$ has a minimum at $\Delta g\simeq 0.7$.
Our experimental results are obtained by the two-terminal conductance
measurement. Therefore the bulk contribution to the channel $\alpha$ becomes maximum at a position displaced from $\Delta g=0.5$ and likely located between
$\Delta g=0.5$ and 1. 
Indeed we find that the bulk contribution is maximal at $\Delta g\simeq 0.7$ which is consistent with measurement results on a Hall-bar device (see SM). 
These results strongly and quantitatively support that channel
$\alpha$ is comprised of the spin-resolved QH bulk state and the
conductance enhancement originates from the equal-spin AR
between the QH bulk state and the superconductor, while the channel
$\beta$ is assigned to normal reflection component in the superconducting bulk gap. We note that
the sub-gap feature and the side peaks have been theoretically predicted for the case of  spin-singlet and spin-triplet superconducting
proximity effect between a topological insulator and a spin-singlet superconductor
with magnetic field~\cite{bursetprb2015}.
There are a few theoretical works describing AR in the QH edge state~\cite{hoppeprl2000,giazottoprb2005,khaymovichel2010,ostaayprb2011},
but none focus on the QH state in the plateau-transition regime, and so further theoretical effort is necessary to reproduce the junction properties between superconductor and spin-resolved QH bulk state.
From the topological aspects, theory predicts that the chiral topological superconductor state can be realized in superconductor-QH state junctions and therefore such junctions can be utilized to study non-Abelian statistics of the Majorana Fermions localized at the center of vortexes near the plateau-transition regime~\cite{xiaoprb2010}. 
Our results indicate that it is possible to induce the proximity gap even in the spin-resolved QH state via equal-spin AR and so realize such chiral topological superconductivity.

In summary, we studied the transport properties of junctions between
a NbTi superconductor and an InAs QW in the spin-resolved
QH regime. 
We observed sub-gap features indicating Andreev transport arising from two channels. One equal spin Andreev reflection channel which produces peaks at zero bias, and a conventional Andreev reflection channel producing side peaks.
These results indicate that junctions of NbTi and the InAs QW are a promising candidate to experimentally study the spin-triplet
superconducting proximity effect in semiconductors and also topological superconductivity.

\clearpage
\section*{Acknowledgment}
We greatly thank Y. Tanaka, P. Burset, and R. S. Deacon for fruitful discussions.
This work was partially supported by Grant-in-Aid for Young Scientific Research (A) (No. JP15H05407), Grant-in-Aid for Scientific Research (A) (No. JP16H02204), Grant-in-Aid for Scientific Research (S) (No. JP26220710), JSPS Research Fellowship for Young Scientists (No. JP14J10600), JSPS Program for Leading Graduate Schools (MERIT) from JSPS, Grant-in-Aid for Scientific Research on Innovative Area, "Nano Spin Conversion Science" (No.JP15H01012), Grant-in-Aid for Scientific Research on Innovative Area, "Topological Materials Science" (Grant No. JP16H00984) from MEXT, CREST, and the Murata Science Foundation.
\clearpage
\section*{Supplemental Material}
\begin{figure}[t]
\centering \includegraphics[width=0.45\linewidth]{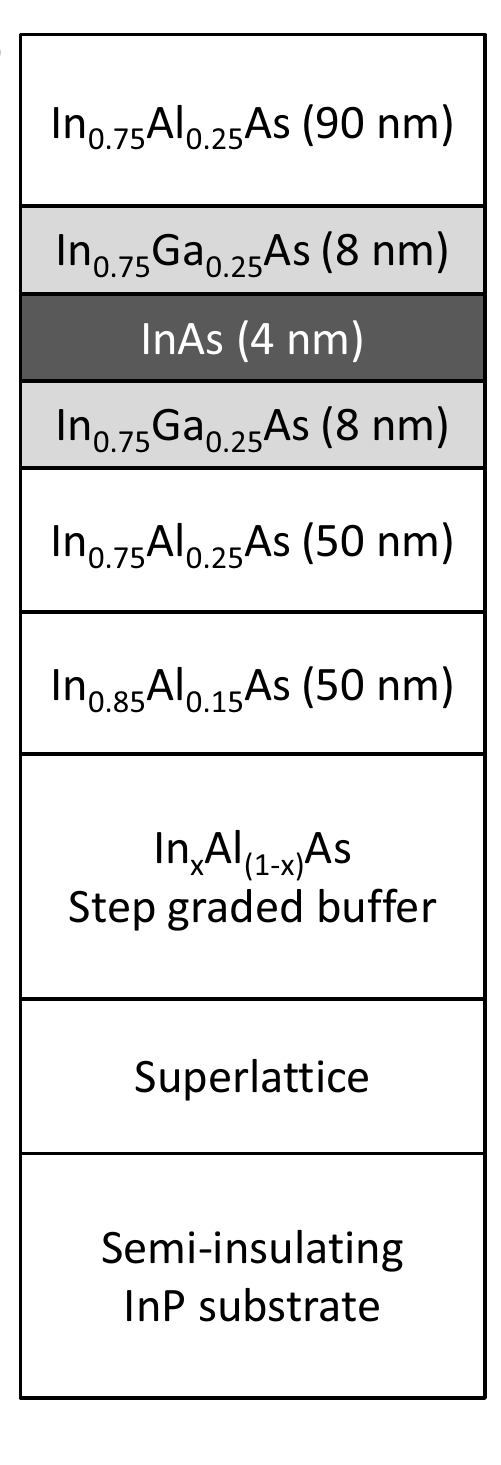} \caption{Schematics of the InAs heterostructure material stack. The2DEG exists in the 4 nm-thick InAs QW.}
\label{stack} 
\end{figure}
\begin{figure}[t]
\centering \includegraphics[width=0.75\linewidth]{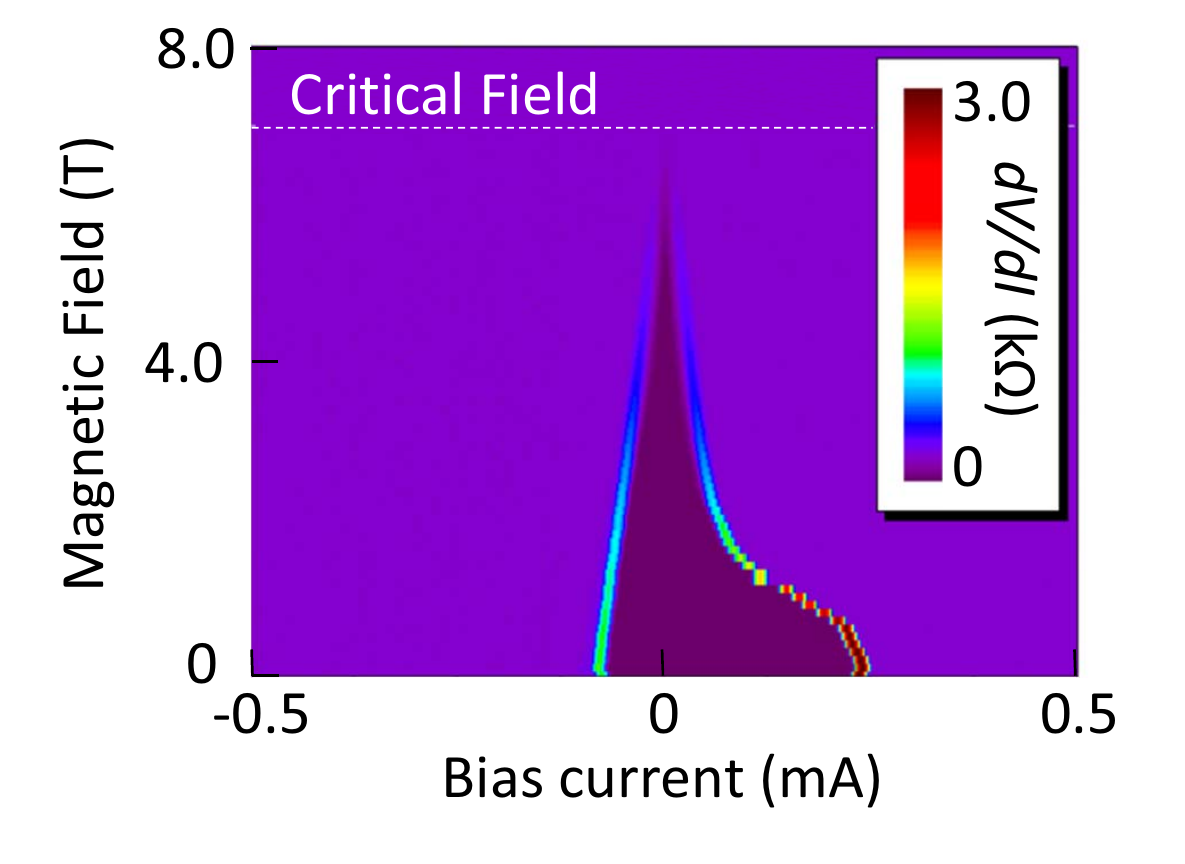} \caption{Measurements of the critical field of a superconducting NbTi 150nm thin film. $dV/dI$ as functions of magnetic field and bias current measured at 2.0 K. The NbTi holds superconductivity when $dV/dI$ is equal
to $0$. The critical field is 7.0 T.}
\label{nbti} 
\end{figure}
\begin{figure}[t]
\centering \includegraphics[width=0.95\linewidth]{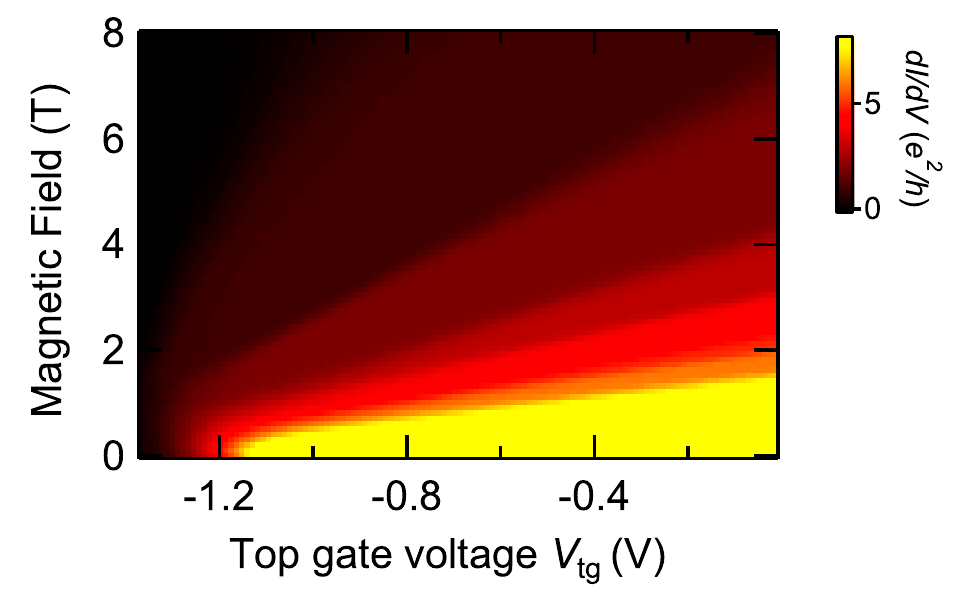} \caption{Conductance as functions of magnetic field and top gate voltage.}
\label{figs2} 
\end{figure}
\begin{figure}[t]
\centering \includegraphics[width=0.95\linewidth]{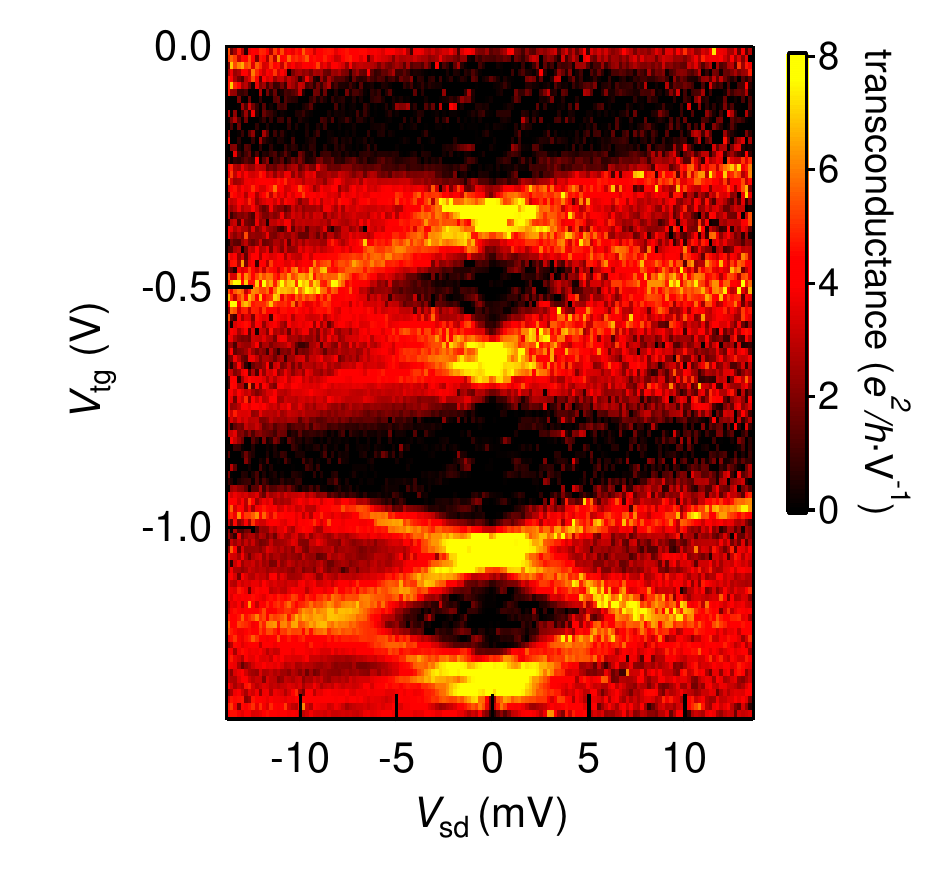} \caption{Transconductance as functions of bias voltage and top gate voltage obtained at 2.4 T.}
\label{figs3} 
\end{figure}
\begin{figure}[t]
\centering \includegraphics[width=0.95\linewidth]{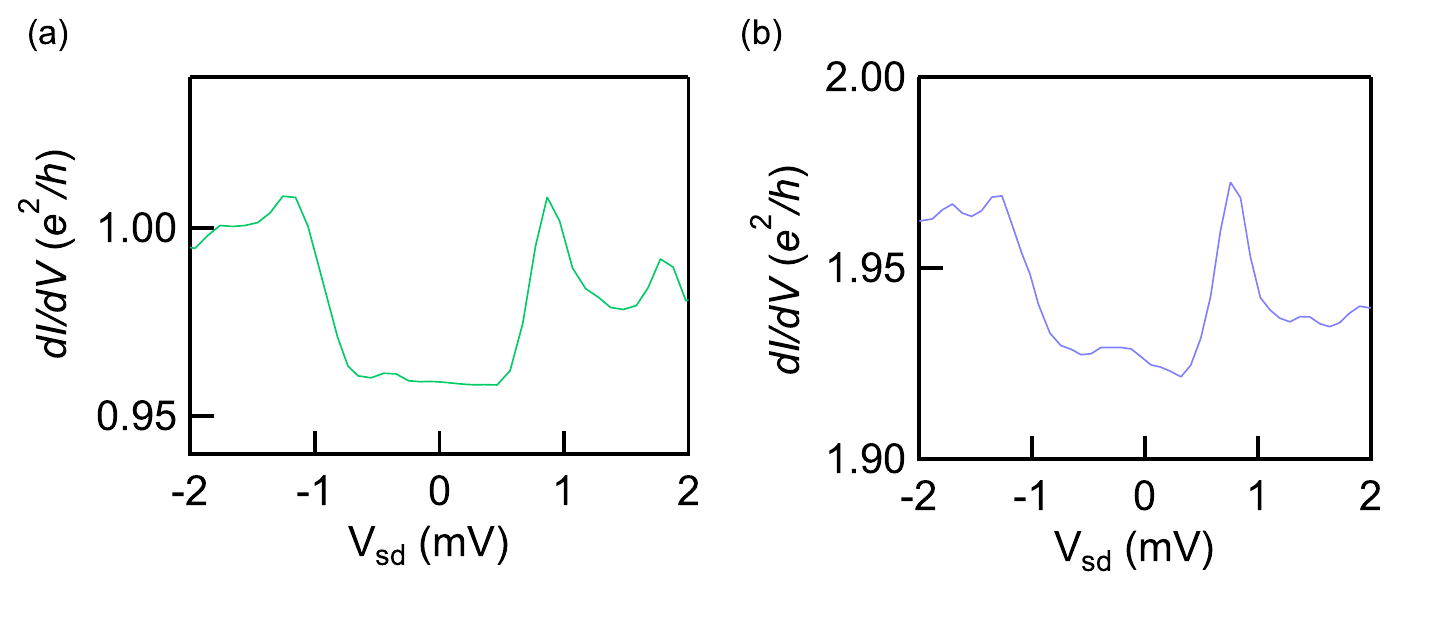} 
\caption{Typical results of the $dI/dV$ vs $V_{\rm sd}$ in the plateau regime obtained at 4 T are shown.  The result on the $\nu = 1$ and 2 plateau is shown in (a) and (b), respectively.}
\label{figs4} 
\end{figure}
\begin{figure}[t]
\centering \includegraphics[width=0.95\linewidth]{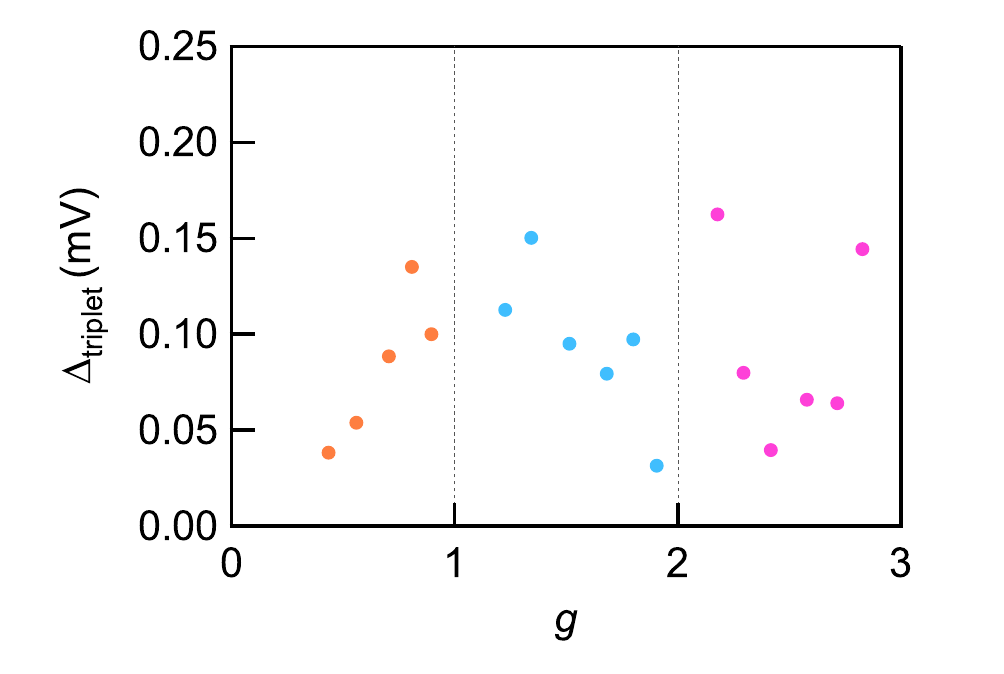} \caption{The calculated spin-triplet superconducting proximity gap energy is
shown. The values are $\sim$0.1 meV.}
\label{figs1} 
\end{figure}
\begin{figure}[t]
\centering \includegraphics[width=0.75\linewidth]{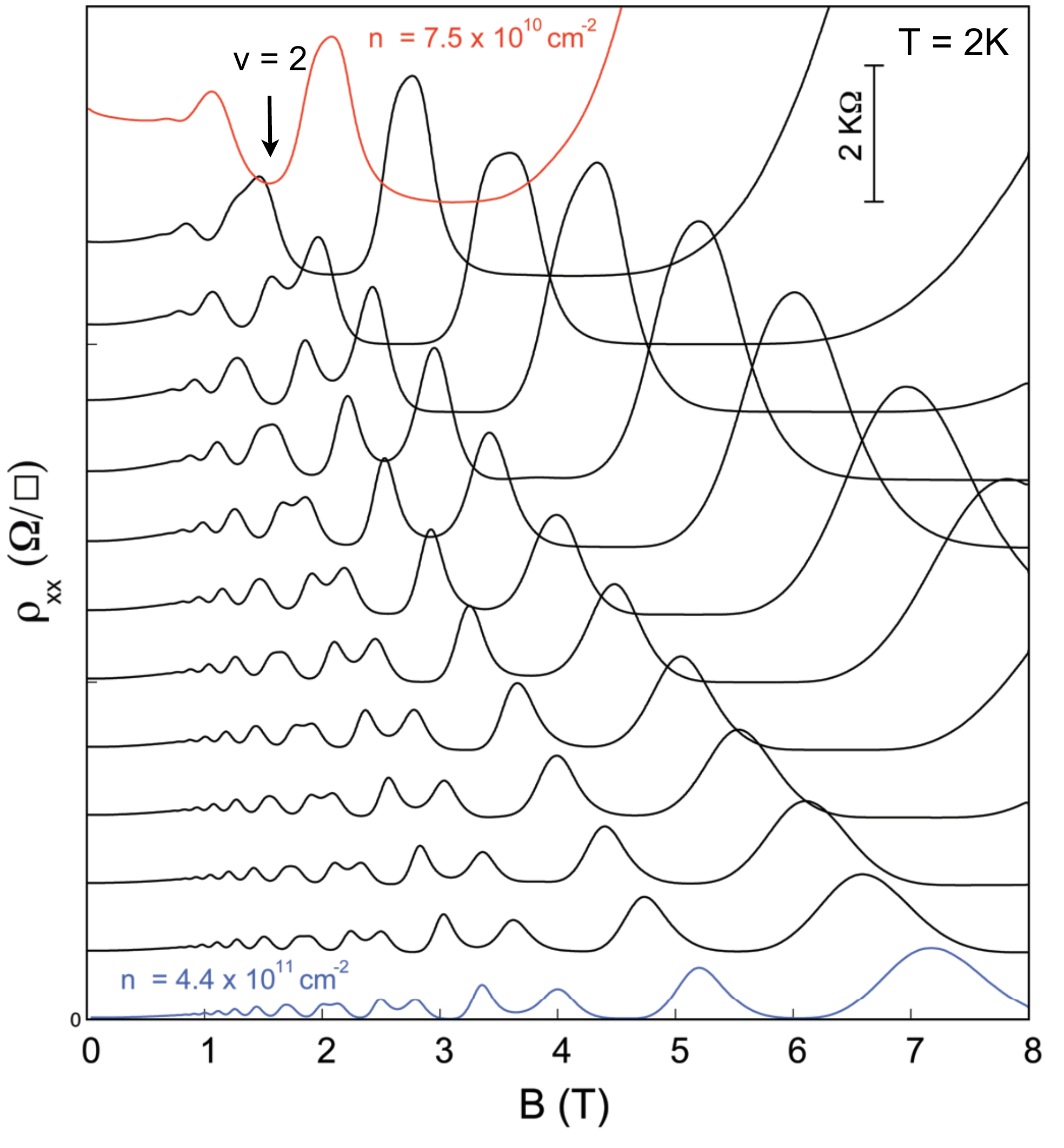} \caption{The sheet resistance as a function of magnetic field is shown. The lines are incrementally shifted for clarity. We
measured the resistance in the various carrier density by tuning the
gate voltage. We subtract the peak height of the resistance around
4 T.}
\label{figsR} 
\end{figure}
\clearpage
\subsection{Device fabrication}
We used an InAs QW grown by molecular beam epitaxy with the density $3\times10^{11}~{\rm cm^{-2}}$
and the mobility $3\times10^{5}~{\rm cm^{2}V^{-1}s^{-1}}$. 
The 2DEG is formed in the 4 nm-thick
InAs layer. The material stack of the InAs heterostructure is schematically shown in Fig.~\ref{stack}.
A mesa was first defined in the substrate by wet etching
with an etchant based on ${\rm H_{3}PO_{4}}$. Then, NbTi was
sputtered to form the superconducting electrodes on the mesa edges,
following a procedure of wet etching to make the clean edge
exposed, sulfur passivation to avoid oxidization of the edge, and
in-situ Ar plasma cleaning. Finally a gate electrode metal
of Titanium and Gold was deposited on top to address the low QH filling
regime even under a low magnetic field. The top gate
is placed on an insulating layer made from cross-linked PMMA. 
This fabrication procedure creates no superconducting
material on the top surface of the mesa.
This is specially devised to control the carrier density not only
of the mesa but also near the junction using the top gate voltage
(discussed later). 
The two junctions are separated by 20 ${\rm \mu}$m, so this device is assumed to have two independent contact regions.

\subsection{NbTi superconductivity}
To characterize the superconducting properties of
the NbTi, we performed a current-bias measurement of the
differential resistance $dV/dI$ at
various out-of-plane magnetic fields $B$ for a 150 nm-thick NbTi thin film device
at 2 K which is lower than the NbTi critical temperature of 6.5
K. The measured data shows a supercurrent branch as $dV/dI$=0 k${\rm \Omega}$ in dark
purple near the zero current bias in Fig.~\ref{nbti}. 
From this data, we evaluate the
critical field of $B=$7.0 T. Herein, the coexistence of superconducting
state and QH states can be realized if the 2DEG is in
the spin-resolved QH regime for $B<7.0$ T.

\subsection{Superconducting proximity at 0 T}
In Fig. 2(a) and (c) in the main text, there are dip structures around $V_{\rm sd}=\pm 0.7$ mV. These dip structures cannot be expected from the normal BTK model. These dip structures have also been reported in experimental studies of junctions of three dimensional topological insulators and superconductor~\cite{snelderarxiv2015}. Some theoretical works predict existence of spin-triplet superconducting proximity on such junctions which generate the dip structures~\cite{bursetprb2014, bursetprb2015}. In our case, strong spin-orbit interaction on the interface can affect the superconducting proximity even at 0 T and invokes finite spin-flip process. Therefore, we suspect that the dip structure may be related to the spin-triplet superconducting proximity even at 0 T.
\subsection{Transconductance in the QH regime}
We measured the conductance as functions of magnetic field and top
gate voltage in order to estimate quality of our InAs quantum Hall
effect and Zeeman effect. Figure \ref{figs2} shows the conductance
as functions of $B$ and $V_{{\rm tg}}$. As seen in this figure,
clear quantized conductance plateaus appear at $B>0.6$~T and the
Zeeman splitting is found at $B>\sim1$~T. 2.4 and 4 T is large enough
to study the coexistence between the spin-resolved QH state and superconductor.
Furthermore, we measured the transconductance defined by the deviation
of $dI/dV$ at 2.4 T. The results are plotted in Fig.~\ref{figs3}.
The diamond-shaped structure can be found. We focus on the bright
areas, namely the plateau-transition regime in the main text. 

\subsection{Sub-gap feature in QH plateau regime}
In the main text, we focus on the sub-gap feature in QH plateau-transition
regime. In this section, we show the measured $dI/dV$ vs $V_{\rm sd}$
in the plateau regime. The typical results obtained at 4 T are shown in Fig.~\ref{figs4}.
Panel (a) and (b) indicates the result on the $\nu = 1$ and 2 plateau, respectively.
Even in the plateau regime, there is a sub-gap feature with a small dip or gap-like structure. These results
are similar to the sub-gap features observed for $\Delta g < 0.4$ of Fig. 3(b) and (c) in the main text. 
The estimated gap from these sub-gap features are nearly equal to the 0.35 mV calculated from the superconducting gap at 0 T. 
This supports our assumption in the main text that the spin-singlet superconducting proximity gap at 4 T is the same as the superconducting gap obtained at 0 T.
We conclude that the QH edge states don't contribute to the Andreev
reflection in the device.

\subsection{Numerical calculation}
We executed the numerical calculation using the model in which we
assume the two channels, $\alpha$ and $\beta$. The fitting parameters are the superconducting gaps $\Delta_{\alpha}$
and $\Delta_{\beta}$, barrier strength $Z_{\alpha}$ ($Z_{\beta}=1$),
normal state interface conductance $G_{n}$, effective temperature
$\omega$, the relative contribution of channel $\alpha$, $P$ and offset
conductance $G_{{\rm offset}}$. We define $G_{{\rm int}}^{\alpha\beta}(V_{{\rm sd}},Z_{\alpha},T,\Delta_{\alpha},\Delta_{\beta})$
as eqn. (2) in the main text. Then the fitting function can be written
as 
\[
\Bigl(\frac{1}{G_{n}\cdot G_{{\rm int}}^{\alpha\beta}(V_{{\rm sd}},Z_{\alpha},T,2\Delta_{\alpha},2\Delta_{\beta})}+\frac{1}{G_{{\rm offset}}}\Bigr)^{-1}.
\]
We take care to confirm that our device has two independent superconducting-QH bulk
state junctions. Therefore, the sub-gap features and the position in $V_{\rm sd}$
of the side peaks are consistent with $2\Delta_{\alpha}$ and $2\Delta_{\beta}$.
To account for the effective temperature, we approximated the deviation of
the Fermi-Dirac distribution function in eqn.~(1) in the main text
as the Gaussian function, $\exp{(-((E-V_{{\rm sd}})/2\omega)^{2})}$,
where $\omega$ is ideally equal to $T$ but now $\omega$ includes
the broadening due to inelastic scattering, inhomogeneity of the gap
and the local heating~\cite{woodsprb2004,degravenl2011}. To execute
the fitting, we constrict the fitting ranges for all the parameters,
and especially we tightly constrict the $\Delta_{\alpha}$ and $\Delta_{\beta}$
from the curve shapes. In order to reproduce the curve shape around
the zero bias voltage, we changed the fitting range for each of the
curves because the differential conductance of the 2DEG appears as
background and the conductance has a large dependence on the bias
voltage near the plateau regime. Due to this background dependence,
we could not reproduce the curve shapes in the two lower curves of the
left panel and middle panel, and the lowest curve of the right panel
in Fig.4(b) of the main text. In these cases, we evaluated only $\Delta_{\alpha}$
and $\Delta_{\beta}$ from the sub-gap peak features ($2\Delta_{\alpha}$ and
$2\Delta_{\beta}$ are indicated on the panels in Fig.4(b) as open
and closed hexagons).
Our fitting scheme includes many free parameters and results are sensitive to the constriction of the variable range. However, the estimated gap energies and $P$ produce relatively constant results with different fitting ranges, so we think it is valuable to discuss these parameters.
All fits are executed with a genetic algorithm (GenCurvefit package for Igor Pro).

\subsection{Calculation of the proximity gap energy}
As written in the main text, we analyzed our experimental data with
the model to evaluate $\Delta_{\alpha}$ and $\Delta_{\beta}$, the superconducting gap energies. However, these values
are enlarged from the true bulk and proximity superconducting gap energies due to
dissipation induced from the bulk state of the mesa. In the plateau
regime, the transport is non-dissipative in the mesa, while the transport
is dissipative in the plateau-transition regime due to the QH bulk
state. Herein, in the plateau-transition regime, applied $V_{{\rm sd}}$
between two superconductors is divided into the voltage on the junctions
and on the mesa, then the deduced $\Delta_{\beta}$ gives a larger
gap energy as the contribution of the QH bulk state in the transport
becomes larger. The equivalent circuit is represented in Fig. 2(d) in the main text.
Consequently, $\Delta_{\beta}$ produces a peak in
the middle of the plateau-transition regime where the bulk contribution
becomes maximum. The true bulk superconducting gap energies (corresponding to the gap
for channel $\beta$), $0.35$ meV, can be evaluated from $\Delta_{\beta}$
near the plateau regime. From this gap energy, we calculated the true
superconducting proximity gap energy, $\sim0.1$ meV as $0.35\times\Delta_{\alpha}/\Delta_{\beta}$.
The calculated gap, $\Delta_{{\rm triplet}}$ as a function
of $g$, $dI/dV$ at $V_{{\rm sd}}=3.5$~mV in unit of $e^{2}/h$
is shown in Fig.\ref{figs1}.

\subsection{The maximum position of the bulk contribution}
Our results for $P$ and $Z_{\alpha}$ have a maximum and minimum,
respectively, at $\Delta g\sim0.7$. In this section, we estimate
how large $\rho_{xx}$ should be to make the bulk contribution have
a maximum at $\Delta g\sim0.7$.

As the shape of our device is square, the two-terminal conductance is
written by $G_{2t}=\sqrt{\sigma_{xx}+\sigma_{xy}}$. $\sigma_{xx}$
and $\sigma_{xy}$ are the longitudinal conductivity and Hall conductivity~\cite{lippmannzna1958, jensenjpc1972, abanin2008},
respectively. $\sigma_{xx}$ has a maximum when the bulk contribution
is maximum and the situation is given by $\sigma_{xy}=0.5e^{2}/h+ne^{2}/h$
($n=0,1,2,3...$), at which change in the filling factor $\Delta\nu$
is equal to 0.5. Herein, if the bulk contribution is maximum at $\Delta g\sim0.7$
(namely $G_{2t}=0.7e^{2}/h+ne^{2}/h$), $\sigma_{xx}(\sigma_{xy})$
should be 0.49$e^{2}/h$ (0.5$e^{2}/h$), 0.80$e^{2}/h$ (1.5$e^{2}/h$),
and 1.02$e^{2}/h$ (2.5$e^{2}/h$) in the $n=0,1,{\rm and}~2$ cases,
respectively. From these conductivity, we can calculate the longitudinal
resistance,$\rho_{xx}=\sigma_{xx}/(\sigma_{xx}^{2}+\sigma_{xy}^{2})$,
resulting in 1.0$h/e^{2}$ ($\sim26{\rm k\Omega}$), 0.28$h/e^{2}$($\sim7.0{\rm k\Omega}$),
and 0.14$h/e^{2}$($\sim3.6{\rm k\Omega}$) in the $n=0,1,{\rm and}~2$
cases, respectively. The calculated longitudinal resistance should
be obtained when the bulk contribution is maximum, meaning $\rho_{xx}$
is maximum with the calculated resistance in the region corresponding
to the plateau-transition regime. We measured a Hall bar device at
2 K, fabricated from the same InAs quantum well wafer as we used for the superconducting devices and the results are shown in Fig.~\ref{figsR}. $\rho_{xx}$
has some peaks consistent with the finite bulk contribution. The maximum
of $\rho_{xx}$ at 4 T is at 4 k${\rm \Omega}$ and 1.5 k${\rm \Omega}$
corresponding to the $n=1{\rm and}~2$ cases, respectively. The measured
resistances are comparable to the estimated resistance based on the
assumption that the bulk contribution is maximum at $\Delta g\sim0.7$
($G_{2t}=0.7e^{2}/h+ne^{2}/h$).

\bibliographystyle{apsrev4-1}

\begin{thebibliography}{47}%
\makeatletter
\providecommand \@ifxundefined [1]{%
 \@ifx{#1\undefined}
}%
\providecommand \@ifnum [1]{%
 \ifnum #1\expandafter \@firstoftwo
 \else \expandafter \@secondoftwo
 \fi
}%
\providecommand \@ifx [1]{%
 \ifx #1\expandafter \@firstoftwo
 \else \expandafter \@secondoftwo
 \fi
}%
\providecommand \natexlab [1]{#1}%
\providecommand \enquote  [1]{``#1''}%
\providecommand \bibnamefont  [1]{#1}%
\providecommand \bibfnamefont [1]{#1}%
\providecommand \citenamefont [1]{#1}%
\providecommand \href@noop [0]{\@secondoftwo}%
\providecommand \href [0]{\begingroup \@sanitize@url \@href}%
\providecommand \@href[1]{\@@startlink{#1}\@@href}%
\providecommand \@@href[1]{\endgroup#1\@@endlink}%
\providecommand \@sanitize@url [0]{\catcode `\\12\catcode `\$12\catcode
  `\&12\catcode `\#12\catcode `\^12\catcode `\_12\catcode `\%12\relax}%
\providecommand \@@startlink[1]{}%
\providecommand \@@endlink[0]{}%
\providecommand \url  [0]{\begingroup\@sanitize@url \@url }%
\providecommand \@url [1]{\endgroup\@href {#1}{\urlprefix }}%
\providecommand \urlprefix  [0]{URL }%
\providecommand \Eprint [0]{\href }%
\providecommand \doibase [0]{http://dx.doi.org/}%
\providecommand \selectlanguage [0]{\@gobble}%
\providecommand \bibinfo  [0]{\@secondoftwo}%
\providecommand \bibfield  [0]{\@secondoftwo}%
\providecommand \translation [1]{[#1]}%
\providecommand \BibitemOpen [0]{}%
\providecommand \bibitemStop [0]{}%
\providecommand \bibitemNoStop [0]{.\EOS\space}%
\providecommand \EOS [0]{\spacefactor3000\relax}%
\providecommand \BibitemShut  [1]{\csname bibitem#1\endcsname}%
\let\auto@bib@innerbib\@empty
\bibitem [{\citenamefont {Andreev}(1964)}]{andreevjetp1964}%
  \BibitemOpen
  \bibfield  {author} {\bibinfo {author} {\bibfnamefont {A.~F.}\ \bibnamefont
  {Andreev}},\ }\href@noop {} {\bibfield  {journal} {\bibinfo  {journal} {Sov.
  Phys. JETP.}\ }\textbf {\bibinfo {volume} {19}},\ \bibinfo {pages} {1228}
  (\bibinfo {year} {1964})}\BibitemShut {NoStop}%
\bibitem [{\citenamefont {Bergeret}\ \emph {et~al.}(2005)\citenamefont
  {Bergeret}, \citenamefont {Volkov},\ and\ \citenamefont
  {Efetov}}]{bergetrmp2005}%
  \BibitemOpen
  \bibfield  {author} {\bibinfo {author} {\bibfnamefont {F.~S.}\ \bibnamefont
  {Bergeret}}, \bibinfo {author} {\bibfnamefont {A.~F.}\ \bibnamefont
  {Volkov}}, \ and\ \bibinfo {author} {\bibfnamefont {K.~B.}\ \bibnamefont
  {Efetov}},\ }\href {\doibase 10.1103/RevModPhys.77.1321} {\bibfield
  {journal} {\bibinfo  {journal} {Rev. Mod. Phys.}\ }\textbf {\bibinfo {volume}
  {77}},\ \bibinfo {pages} {1321} (\bibinfo {year} {2005})}\BibitemShut
  {NoStop}%
\bibitem [{\citenamefont {Keizer}\ \emph {et~al.}(2010)\citenamefont {Keizer},
  \citenamefont {Goennenwein}, \citenamefont {Klapwijk}, \citenamefont {Miao},
  \citenamefont {Xiao},\ and\ \citenamefont {Gupta}}]{keizernature2006}%
  \BibitemOpen
  \bibfield  {author} {\bibinfo {author} {\bibfnamefont {R.~S.}\ \bibnamefont
  {Keizer}}, \bibinfo {author} {\bibfnamefont {S.~T.~B.}\ \bibnamefont
  {Goennenwein}}, \bibinfo {author} {\bibfnamefont {T.~M.}\ \bibnamefont
  {Klapwijk}}, \bibinfo {author} {\bibfnamefont {G.}~\bibnamefont {Miao}},
  \bibinfo {author} {\bibfnamefont {G.}~\bibnamefont {Xiao}}, \ and\ \bibinfo
  {author} {\bibfnamefont {A.}~\bibnamefont {Gupta}},\ }\href@noop {}
  {\bibfield  {journal} {\bibinfo  {journal} {Nature}\ }\textbf {\bibinfo
  {volume} {439}},\ \bibinfo {pages} {825} (\bibinfo {year}
  {2010})}\BibitemShut {NoStop}%
\bibitem [{\citenamefont {Asano}\ \emph
  {et~al.}(2007{\natexlab{a}})\citenamefont {Asano}, \citenamefont {Tanaka},\
  and\ \citenamefont {Golubov}}]{asanoprl2007}%
  \BibitemOpen
  \bibfield  {author} {\bibinfo {author} {\bibfnamefont {Y.}~\bibnamefont
  {Asano}}, \bibinfo {author} {\bibfnamefont {Y.}~\bibnamefont {Tanaka}}, \
  and\ \bibinfo {author} {\bibfnamefont {A.~A.}\ \bibnamefont {Golubov}},\
  }\href {\doibase 10.1103/PhysRevLett.98.107002} {\bibfield  {journal}
  {\bibinfo  {journal} {Phys. Rev. Lett.}\ }\textbf {\bibinfo {volume} {98}},\
  \bibinfo {pages} {107002} (\bibinfo {year} {2007}{\natexlab{a}})}\BibitemShut
  {NoStop}%
\bibitem [{\citenamefont {Asano}\ \emph
  {et~al.}(2007{\natexlab{b}})\citenamefont {Asano}, \citenamefont {Sawa},
  \citenamefont {Tanaka},\ and\ \citenamefont {Golubov}}]{asanoprb2007}%
  \BibitemOpen
  \bibfield  {author} {\bibinfo {author} {\bibfnamefont {Y.}~\bibnamefont
  {Asano}}, \bibinfo {author} {\bibfnamefont {Y.}~\bibnamefont {Sawa}},
  \bibinfo {author} {\bibfnamefont {Y.}~\bibnamefont {Tanaka}}, \ and\ \bibinfo
  {author} {\bibfnamefont {A.~A.}\ \bibnamefont {Golubov}},\ }\href {\doibase
  10.1103/PhysRevB.76.224525} {\bibfield  {journal} {\bibinfo  {journal} {Phys.
  Rev. B}\ }\textbf {\bibinfo {volume} {76}},\ \bibinfo {pages} {224525}
  (\bibinfo {year} {2007}{\natexlab{b}})}\BibitemShut {NoStop}%
\bibitem [{\citenamefont {Robinson}\ \emph {et~al.}(2010)\citenamefont
  {Robinson}, \citenamefont {Witt},\ and\ \citenamefont
  {Blamire}}]{robinsonscience2010}%
  \BibitemOpen
  \bibfield  {author} {\bibinfo {author} {\bibfnamefont {J.~W.~A.}\
  \bibnamefont {Robinson}}, \bibinfo {author} {\bibfnamefont {J.~D.~S.}\
  \bibnamefont {Witt}}, \ and\ \bibinfo {author} {\bibfnamefont {M.~G.}\
  \bibnamefont {Blamire}},\ }\href {\doibase 10.1126/science.1189246}
  {\bibfield  {journal} {\bibinfo  {journal} {Science}\ }\textbf {\bibinfo
  {volume} {329}},\ \bibinfo {pages} {59} (\bibinfo {year} {2010})}\BibitemShut
  {NoStop}%
\bibitem [{\citenamefont {Visani}\ \emph {et~al.}(2012)\citenamefont {Visani},
  \citenamefont {Sefrioui}, \citenamefont {Tornos}, \citenamefont {Leon},
  \citenamefont {Briatico}, \citenamefont {Bibes}, \citenamefont {Barthelemy},
  \citenamefont {Santamaria},\ and\ \citenamefont
  {Villegas}}]{visaninatphys2012}%
  \BibitemOpen
  \bibfield  {author} {\bibinfo {author} {\bibfnamefont {C.}~\bibnamefont
  {Visani}}, \bibinfo {author} {\bibfnamefont {Z.}~\bibnamefont {Sefrioui}},
  \bibinfo {author} {\bibfnamefont {J.}~\bibnamefont {Tornos}}, \bibinfo
  {author} {\bibfnamefont {C.}~\bibnamefont {Leon}}, \bibinfo {author}
  {\bibfnamefont {J.}~\bibnamefont {Briatico}}, \bibinfo {author}
  {\bibfnamefont {M.}~\bibnamefont {Bibes}}, \bibinfo {author} {\bibfnamefont
  {A.}~\bibnamefont {Barthelemy}}, \bibinfo {author} {\bibfnamefont
  {J.}~\bibnamefont {Santamaria}}, \ and\ \bibinfo {author} {\bibfnamefont
  {J.~E.}\ \bibnamefont {Villegas}},\ }\href@noop {} {\bibfield  {journal}
  {\bibinfo  {journal} {Nature Physics}\ }\textbf {\bibinfo {volume} {8}},\
  \bibinfo {pages} {539} (\bibinfo {year} {2012})}\BibitemShut {NoStop}%
\bibitem [{\citenamefont {Linder}\ and\ \citenamefont
  {Robinson}(2015)}]{lindernatphys2015}%
  \BibitemOpen
  \bibfield  {author} {\bibinfo {author} {\bibfnamefont {J.}~\bibnamefont
  {Linder}}\ and\ \bibinfo {author} {\bibfnamefont {J.~W.~A.}\ \bibnamefont
  {Robinson}},\ }\href {\doibase 10.1038/nphys3242} {\bibfield  {journal}
  {\bibinfo  {journal} {Nature Physics}\ }\textbf {\bibinfo {volume} {11}},\
  \bibinfo {pages} {307} (\bibinfo {year} {2015})}\BibitemShut {NoStop}%
\bibitem [{\citenamefont {Eschrig}\ and\ \citenamefont
  {Lofwander}(2008)}]{eschrignatphys2008}%
  \BibitemOpen
  \bibfield  {author} {\bibinfo {author} {\bibfnamefont {M.}~\bibnamefont
  {Eschrig}}\ and\ \bibinfo {author} {\bibfnamefont {T.}~\bibnamefont
  {Lofwander}},\ }\href {\doibase 10.1038/nphys831} {\bibfield  {journal}
  {\bibinfo  {journal} {Nature Physics}\ }\textbf {\bibinfo {volume} {4}},\
  \bibinfo {pages} {138} (\bibinfo {year} {2008})}\BibitemShut {NoStop}%
\bibitem [{\citenamefont {van Ostaay}\ \emph {et~al.}(2011)\citenamefont {van
  Ostaay}, \citenamefont {Akhmerov},\ and\ \citenamefont
  {Beenakker}}]{ostaayprb2011}%
  \BibitemOpen
  \bibfield  {author} {\bibinfo {author} {\bibfnamefont {J.~A.~M.}\
  \bibnamefont {van Ostaay}}, \bibinfo {author} {\bibfnamefont {A.~R.}\
  \bibnamefont {Akhmerov}}, \ and\ \bibinfo {author} {\bibfnamefont {C.~W.~J.}\
  \bibnamefont {Beenakker}},\ }\href {\doibase 10.1103/PhysRevB.83.195441}
  {\bibfield  {journal} {\bibinfo  {journal} {Phys. Rev. B}\ }\textbf {\bibinfo
  {volume} {83}},\ \bibinfo {pages} {195441} (\bibinfo {year}
  {2011})}\BibitemShut {NoStop}%
\bibitem [{\citenamefont {Luo}\ \emph {et~al.}(1988)\citenamefont {Luo},
  \citenamefont {Munekata}, \citenamefont {Fang},\ and\ \citenamefont
  {Stiles}}]{luoprb1988}%
  \BibitemOpen
  \bibfield  {author} {\bibinfo {author} {\bibfnamefont {J.}~\bibnamefont
  {Luo}}, \bibinfo {author} {\bibfnamefont {H.}~\bibnamefont {Munekata}},
  \bibinfo {author} {\bibfnamefont {F.~F.}\ \bibnamefont {Fang}}, \ and\
  \bibinfo {author} {\bibfnamefont {P.~J.}\ \bibnamefont {Stiles}},\ }\href
  {\doibase 10.1103/PhysRevB.38.10142} {\bibfield  {journal} {\bibinfo
  {journal} {Phys. Rev. B}\ }\textbf {\bibinfo {volume} {38}},\ \bibinfo
  {pages} {10142} (\bibinfo {year} {1988})}\BibitemShut {NoStop}%
\bibitem [{\citenamefont {de~Andrada~e Silva}\ \emph
  {et~al.}(1997)\citenamefont {de~Andrada~e Silva}, \citenamefont {La~Rocca},\
  and\ \citenamefont {Bassani}}]{silvaprb1997}%
  \BibitemOpen
  \bibfield  {author} {\bibinfo {author} {\bibfnamefont {E.~A.}\ \bibnamefont
  {de~Andrada~e Silva}}, \bibinfo {author} {\bibfnamefont {G.~C.}\ \bibnamefont
  {La~Rocca}}, \ and\ \bibinfo {author} {\bibfnamefont {F.}~\bibnamefont
  {Bassani}},\ }\href {\doibase 10.1103/PhysRevB.55.16293} {\bibfield
  {journal} {\bibinfo  {journal} {Phys. Rev. B}\ }\textbf {\bibinfo {volume}
  {55}},\ \bibinfo {pages} {16293} (\bibinfo {year} {1997})}\BibitemShut
  {NoStop}%
\bibitem [{\citenamefont {Nitta}\ \emph {et~al.}(1997)\citenamefont {Nitta},
  \citenamefont {Akazaki}, \citenamefont {Takayanagi},\ and\ \citenamefont
  {Enoki}}]{nittaprl1997}%
  \BibitemOpen
  \bibfield  {author} {\bibinfo {author} {\bibfnamefont {J.}~\bibnamefont
  {Nitta}}, \bibinfo {author} {\bibfnamefont {T.}~\bibnamefont {Akazaki}},
  \bibinfo {author} {\bibfnamefont {H.}~\bibnamefont {Takayanagi}}, \ and\
  \bibinfo {author} {\bibfnamefont {T.}~\bibnamefont {Enoki}},\ }\href
  {\doibase 10.1103/PhysRevLett.78.1335} {\bibfield  {journal} {\bibinfo
  {journal} {Phys. Rev. Lett.}\ }\textbf {\bibinfo {volume} {78}},\ \bibinfo
  {pages} {1335} (\bibinfo {year} {1997})}\BibitemShut {NoStop}%
\bibitem [{\citenamefont {Heida}\ \emph {et~al.}(1998)\citenamefont {Heida},
  \citenamefont {van Wees}, \citenamefont {Kuipers}, \citenamefont {Klapwijk},\
  and\ \citenamefont {Borghs}}]{heidaprb1998}%
  \BibitemOpen
  \bibfield  {author} {\bibinfo {author} {\bibfnamefont {J.~P.}\ \bibnamefont
  {Heida}}, \bibinfo {author} {\bibfnamefont {B.~J.}\ \bibnamefont {van Wees}},
  \bibinfo {author} {\bibfnamefont {J.~J.}\ \bibnamefont {Kuipers}}, \bibinfo
  {author} {\bibfnamefont {T.~M.}\ \bibnamefont {Klapwijk}}, \ and\ \bibinfo
  {author} {\bibfnamefont {G.}~\bibnamefont {Borghs}},\ }\href {\doibase
  10.1103/PhysRevB.57.11911} {\bibfield  {journal} {\bibinfo  {journal} {Phys.
  Rev. B}\ }\textbf {\bibinfo {volume} {57}},\ \bibinfo {pages} {11911}
  (\bibinfo {year} {1998})}\BibitemShut {NoStop}%
\bibitem [{\citenamefont {Takayanagi}\ and\ \citenamefont
  {Akazaki}(1998)}]{takayanagiphysb1998}%
  \BibitemOpen
  \bibfield  {author} {\bibinfo {author} {\bibfnamefont {H.}~\bibnamefont
  {Takayanagi}}\ and\ \bibinfo {author} {\bibfnamefont {T.}~\bibnamefont
  {Akazaki}},\ }\href {\doibase
  http://dx.doi.org/10.1016/S0921-4526(98)00164-1} {\bibfield  {journal}
  {\bibinfo  {journal} {Physica B: Condensed Matter}\ }\textbf {\bibinfo
  {volume} {249-251}},\ \bibinfo {pages} {462} (\bibinfo {year}
  {1998})}\BibitemShut {NoStop}%
\bibitem [{\citenamefont {Eroms}\ \emph {et~al.}(2005)\citenamefont {Eroms},
  \citenamefont {Weiss}, \citenamefont {Boeck}, \citenamefont {Borghs},\ and\
  \citenamefont {Z\"ulicke}}]{eromsprl2005}%
  \BibitemOpen
  \bibfield  {author} {\bibinfo {author} {\bibfnamefont {J.}~\bibnamefont
  {Eroms}}, \bibinfo {author} {\bibfnamefont {D.}~\bibnamefont {Weiss}},
  \bibinfo {author} {\bibfnamefont {J.~D.}\ \bibnamefont {Boeck}}, \bibinfo
  {author} {\bibfnamefont {G.}~\bibnamefont {Borghs}}, \ and\ \bibinfo {author}
  {\bibfnamefont {U.}~\bibnamefont {Z\"ulicke}},\ }\href {\doibase
  10.1103/PhysRevLett.95.107001} {\bibfield  {journal} {\bibinfo  {journal}
  {Phys. Rev. Lett.}\ }\textbf {\bibinfo {volume} {95}},\ \bibinfo {pages}
  {107001} (\bibinfo {year} {2005})}\BibitemShut {NoStop}%
\bibitem [{\citenamefont {Rickhaus}\ \emph {et~al.}(2012)\citenamefont
  {Rickhaus}, \citenamefont {Weiss}, \citenamefont {Marot},\ and\ \citenamefont
  {Schonenberger}}]{rickhausnl2012}%
  \BibitemOpen
  \bibfield  {author} {\bibinfo {author} {\bibfnamefont {P.}~\bibnamefont
  {Rickhaus}}, \bibinfo {author} {\bibfnamefont {M.}~\bibnamefont {Weiss}},
  \bibinfo {author} {\bibfnamefont {L.}~\bibnamefont {Marot}}, \ and\ \bibinfo
  {author} {\bibfnamefont {C.}~\bibnamefont {Schonenberger}},\ }\href {\doibase
  10.1021/nl204415s} {\bibfield  {journal} {\bibinfo  {journal} {Nano Letters}\
  }\textbf {\bibinfo {volume} {12}},\ \bibinfo {pages} {1942} (\bibinfo {year}
  {2012})},\ \bibinfo {note} {pMID: 22417183}\BibitemShut {NoStop}%
\bibitem [{\citenamefont {Komatsu}\ \emph {et~al.}(2012)\citenamefont
  {Komatsu}, \citenamefont {Li}, \citenamefont {Autier-Laurent}, \citenamefont
  {Bouchiat},\ and\ \citenamefont {Gu\'eron}}]{komatsuprb2012}%
  \BibitemOpen
  \bibfield  {author} {\bibinfo {author} {\bibfnamefont {K.}~\bibnamefont
  {Komatsu}}, \bibinfo {author} {\bibfnamefont {C.}~\bibnamefont {Li}},
  \bibinfo {author} {\bibfnamefont {S.}~\bibnamefont {Autier-Laurent}},
  \bibinfo {author} {\bibfnamefont {H.}~\bibnamefont {Bouchiat}}, \ and\
  \bibinfo {author} {\bibfnamefont {S.}~\bibnamefont {Gu\'eron}},\ }\href
  {\doibase 10.1103/PhysRevB.86.115412} {\bibfield  {journal} {\bibinfo
  {journal} {Phys. Rev. B}\ }\textbf {\bibinfo {volume} {86}},\ \bibinfo
  {pages} {115412} (\bibinfo {year} {2012})}\BibitemShut {NoStop}%
\bibitem [{\citenamefont {Wan}\ \emph {et~al.}(2015)\citenamefont {Wan},
  \citenamefont {Kazakov}, \citenamefont {Manfra}, \citenamefont {Pfeiffer},
  \citenamefont {West},\ and\ \citenamefont {Rokhinson}}]{wannatcommun2015}%
  \BibitemOpen
  \bibfield  {author} {\bibinfo {author} {\bibfnamefont {Z.}~\bibnamefont
  {Wan}}, \bibinfo {author} {\bibfnamefont {A.}~\bibnamefont {Kazakov}},
  \bibinfo {author} {\bibfnamefont {M.~J.}\ \bibnamefont {Manfra}}, \bibinfo
  {author} {\bibfnamefont {L.~N.}\ \bibnamefont {Pfeiffer}}, \bibinfo {author}
  {\bibfnamefont {K.~W.}\ \bibnamefont {West}}, \ and\ \bibinfo {author}
  {\bibfnamefont {L.~P.}\ \bibnamefont {Rokhinson}},\ }\href {\doibase
  http://dx.doi.org/10.1038/ncomms8426} {\bibfield  {journal} {\bibinfo
  {journal} {Nature Communications}\ }\textbf {\bibinfo {volume} {6}},\
  \bibinfo {pages} {7426} (\bibinfo {year} {2015})}\BibitemShut {NoStop}%
\bibitem [{\citenamefont {Amet}\ \emph {et~al.}(2016)\citenamefont {Amet},
  \citenamefont {Ke}, \citenamefont {Borzenets}, \citenamefont {Wang},
  \citenamefont {Watanabe}, \citenamefont {Taniguchi}, \citenamefont {Deacon},
  \citenamefont {Yamamoto}, \citenamefont {Bomze}, \citenamefont {Tarucha},\
  and\ \citenamefont {Finkelstein}}]{ametscience2016}%
  \BibitemOpen
  \bibfield  {author} {\bibinfo {author} {\bibfnamefont {F.}~\bibnamefont
  {Amet}}, \bibinfo {author} {\bibfnamefont {C.~T.}\ \bibnamefont {Ke}},
  \bibinfo {author} {\bibfnamefont {I.~V.}\ \bibnamefont {Borzenets}}, \bibinfo
  {author} {\bibfnamefont {J.}~\bibnamefont {Wang}}, \bibinfo {author}
  {\bibfnamefont {K.}~\bibnamefont {Watanabe}}, \bibinfo {author}
  {\bibfnamefont {T.}~\bibnamefont {Taniguchi}}, \bibinfo {author}
  {\bibfnamefont {R.~S.}\ \bibnamefont {Deacon}}, \bibinfo {author}
  {\bibfnamefont {M.}~\bibnamefont {Yamamoto}}, \bibinfo {author}
  {\bibfnamefont {Y.}~\bibnamefont {Bomze}}, \bibinfo {author} {\bibfnamefont
  {S.}~\bibnamefont {Tarucha}}, \ and\ \bibinfo {author} {\bibfnamefont
  {G.}~\bibnamefont {Finkelstein}},\ }\href {\doibase 10.1126/science.aad6203}
  {\bibfield  {journal} {\bibinfo  {journal} {Science}\ }\textbf {\bibinfo
  {volume} {352}},\ \bibinfo {pages} {966} (\bibinfo {year}
  {2016})}\BibitemShut {NoStop}%
\bibitem [{\citenamefont {Qi}\ \emph {et~al.}(2010)\citenamefont {Qi},
  \citenamefont {Hughes},\ and\ \citenamefont {Zhang}}]{xiaoprb2010}%
  \BibitemOpen
  \bibfield  {author} {\bibinfo {author} {\bibfnamefont {X.-L.}\ \bibnamefont
  {Qi}}, \bibinfo {author} {\bibfnamefont {T.~L.}\ \bibnamefont {Hughes}}, \
  and\ \bibinfo {author} {\bibfnamefont {S.-C.}\ \bibnamefont {Zhang}},\ }\href
  {\doibase 10.1103/PhysRevB.82.184516} {\bibfield  {journal} {\bibinfo
  {journal} {Phys. Rev. B}\ }\textbf {\bibinfo {volume} {82}},\ \bibinfo
  {pages} {184516} (\bibinfo {year} {2010})}\BibitemShut {NoStop}%
\bibitem [{\citenamefont {Fu}\ and\ \citenamefont {Kane}(2008)}]{fuprl2008}%
  \BibitemOpen
  \bibfield  {author} {\bibinfo {author} {\bibfnamefont {L.}~\bibnamefont
  {Fu}}\ and\ \bibinfo {author} {\bibfnamefont {C.~L.}\ \bibnamefont {Kane}},\
  }\href {\doibase 10.1103/PhysRevLett.100.096407} {\bibfield  {journal}
  {\bibinfo  {journal} {Phys. Rev. Lett.}\ }\textbf {\bibinfo {volume} {100}},\
  \bibinfo {pages} {096407} (\bibinfo {year} {2008})}\BibitemShut {NoStop}%
\bibitem [{\citenamefont {Hasan}\ and\ \citenamefont
  {Kane}(2010)}]{hasanrmp2010}%
  \BibitemOpen
  \bibfield  {author} {\bibinfo {author} {\bibfnamefont {M.~Z.}\ \bibnamefont
  {Hasan}}\ and\ \bibinfo {author} {\bibfnamefont {C.~L.}\ \bibnamefont
  {Kane}},\ }\href {\doibase 10.1103/RevModPhys.82.3045} {\bibfield  {journal}
  {\bibinfo  {journal} {Rev. Mod. Phys.}\ }\textbf {\bibinfo {volume} {82}},\
  \bibinfo {pages} {3045} (\bibinfo {year} {2010})}\BibitemShut {NoStop}%
\bibitem [{\citenamefont {Mourik}\ \emph {et~al.}(2012)\citenamefont {Mourik},
  \citenamefont {Zuo}, \citenamefont {Frolov}, \citenamefont {Plissard},
  \citenamefont {Bakkers},\ and\ \citenamefont
  {Kouwenhoven}}]{mourikscience2012}%
  \BibitemOpen
  \bibfield  {author} {\bibinfo {author} {\bibfnamefont {V.}~\bibnamefont
  {Mourik}}, \bibinfo {author} {\bibfnamefont {K.}~\bibnamefont {Zuo}},
  \bibinfo {author} {\bibfnamefont {S.~M.}\ \bibnamefont {Frolov}}, \bibinfo
  {author} {\bibfnamefont {S.~R.}\ \bibnamefont {Plissard}}, \bibinfo {author}
  {\bibfnamefont {E.~P. A.~M.}\ \bibnamefont {Bakkers}}, \ and\ \bibinfo
  {author} {\bibfnamefont {L.~P.}\ \bibnamefont {Kouwenhoven}},\ }\href
  {\doibase 10.1126/science.1222360} {\bibfield  {journal} {\bibinfo  {journal}
  {Science}\ }\textbf {\bibinfo {volume} {336}},\ \bibinfo {pages} {1003}
  (\bibinfo {year} {2012})}\BibitemShut {NoStop}%
\bibitem [{\citenamefont {Das}\ \emph {et~al.}(2012)\citenamefont {Das},
  \citenamefont {Ronen}, \citenamefont {Most}, \citenamefont {Oreg},
  \citenamefont {Heiblum},\ and\ \citenamefont {Shtrikman}}]{dasnatphys2012}%
  \BibitemOpen
  \bibfield  {author} {\bibinfo {author} {\bibfnamefont {A.}~\bibnamefont
  {Das}}, \bibinfo {author} {\bibfnamefont {Y.}~\bibnamefont {Ronen}}, \bibinfo
  {author} {\bibfnamefont {Y.}~\bibnamefont {Most}}, \bibinfo {author}
  {\bibfnamefont {Y.}~\bibnamefont {Oreg}}, \bibinfo {author} {\bibfnamefont
  {M.}~\bibnamefont {Heiblum}}, \ and\ \bibinfo {author} {\bibfnamefont
  {H.}~\bibnamefont {Shtrikman}},\ }\href@noop {} {\bibfield  {journal}
  {\bibinfo  {journal} {Nature Physics}\ }\textbf {\bibinfo {volume} {8}},\
  \bibinfo {pages} {887} (\bibinfo {year} {2012})}\BibitemShut {NoStop}%
\bibitem [{\citenamefont {Rokhinson}\ \emph {et~al.}(2012)\citenamefont
  {Rokhinson}, \citenamefont {Liu},\ and\ \citenamefont
  {Furdyna}}]{rokhinsonnatphys2012}%
  \BibitemOpen
  \bibfield  {author} {\bibinfo {author} {\bibfnamefont {L.~P.}\ \bibnamefont
  {Rokhinson}}, \bibinfo {author} {\bibfnamefont {X.}~\bibnamefont {Liu}}, \
  and\ \bibinfo {author} {\bibfnamefont {J.~K.}\ \bibnamefont {Furdyna}},\
  }\href@noop {} {\bibfield  {journal} {\bibinfo  {journal} {Nature Physics}\
  }\textbf {\bibinfo {volume} {8}},\ \bibinfo {pages} {795} (\bibinfo {year}
  {2012})}\BibitemShut {NoStop}%
\bibitem [{\citenamefont {Nadj-Perge}\ \emph {et~al.}(2014)\citenamefont
  {Nadj-Perge}, \citenamefont {Drozdov}, \citenamefont {Li}, \citenamefont
  {Chen}, \citenamefont {Jeon}, \citenamefont {Seo}, \citenamefont {MacDonald},
  \citenamefont {Bernevig},\ and\ \citenamefont
  {Yazdani}}]{nadjpergescience2014}%
  \BibitemOpen
  \bibfield  {author} {\bibinfo {author} {\bibfnamefont {S.}~\bibnamefont
  {Nadj-Perge}}, \bibinfo {author} {\bibfnamefont {I.~K.}\ \bibnamefont
  {Drozdov}}, \bibinfo {author} {\bibfnamefont {J.}~\bibnamefont {Li}},
  \bibinfo {author} {\bibfnamefont {H.}~\bibnamefont {Chen}}, \bibinfo {author}
  {\bibfnamefont {S.}~\bibnamefont {Jeon}}, \bibinfo {author} {\bibfnamefont
  {J.}~\bibnamefont {Seo}}, \bibinfo {author} {\bibfnamefont {A.~H.}\
  \bibnamefont {MacDonald}}, \bibinfo {author} {\bibfnamefont {B.~A.}\
  \bibnamefont {Bernevig}}, \ and\ \bibinfo {author} {\bibfnamefont
  {A.}~\bibnamefont {Yazdani}},\ }\href {\doibase 10.1126/science.1259327}
  {\bibfield  {journal} {\bibinfo  {journal} {Science}\ }\textbf {\bibinfo
  {volume} {346}},\ \bibinfo {pages} {602} (\bibinfo {year}
  {2014})}\BibitemShut {NoStop}%
\bibitem [{\citenamefont {Bocquillon}\ \emph {et~al.}(2016)\citenamefont
  {Bocquillon}, \citenamefont {Deacon}, \citenamefont {Wiedenmann},
  \citenamefont {Leubner}, \citenamefont {Klapwijk}, \citenamefont {Brune},
  \citenamefont {Ishibashi}, \citenamefont {Buhmann},\ and\ \citenamefont
  {Molenkamp}}]{boucquillonnatnano2016}%
  \BibitemOpen
  \bibfield  {author} {\bibinfo {author} {\bibfnamefont {E.}~\bibnamefont
  {Bocquillon}}, \bibinfo {author} {\bibfnamefont {R.~S.}\ \bibnamefont
  {Deacon}}, \bibinfo {author} {\bibfnamefont {J.}~\bibnamefont {Wiedenmann}},
  \bibinfo {author} {\bibfnamefont {P.}~\bibnamefont {Leubner}}, \bibinfo
  {author} {\bibfnamefont {T.~M.}\ \bibnamefont {Klapwijk}}, \bibinfo {author}
  {\bibfnamefont {C.}~\bibnamefont {Brune}}, \bibinfo {author} {\bibfnamefont
  {K.}~\bibnamefont {Ishibashi}}, \bibinfo {author} {\bibfnamefont
  {H.}~\bibnamefont {Buhmann}}, \ and\ \bibinfo {author} {\bibfnamefont
  {L.~W.}\ \bibnamefont {Molenkamp}},\ }\href@noop {} {\bibfield  {journal}
  {\bibinfo  {journal} {Nature Nanotechnology}\ }\textbf {\bibinfo {volume}
  {12}},\ \bibinfo {pages} {137} (\bibinfo {year} {2016})}\BibitemShut
  {NoStop}%
\bibitem [{\citenamefont {Wiedenmann}\ \emph {et~al.}(2016)\citenamefont
  {Wiedenmann}, \citenamefont {Bocquillon}, \citenamefont {Deacon},
  \citenamefont {Hartinger}, \citenamefont {Herrmann}, \citenamefont
  {Klapwijk}, \citenamefont {Maier}, \citenamefont {Ames}, \citenamefont
  {Brune}, \citenamefont {Gould}, \citenamefont {Oiwa}, \citenamefont
  {Ishibashi}, \citenamefont {Tarucha}, \citenamefont {Buhmann},\ and\
  \citenamefont {Molenkamp}}]{wiedenmannnatcommun2016}%
  \BibitemOpen
  \bibfield  {author} {\bibinfo {author} {\bibfnamefont {J.}~\bibnamefont
  {Wiedenmann}}, \bibinfo {author} {\bibfnamefont {E.}~\bibnamefont
  {Bocquillon}}, \bibinfo {author} {\bibfnamefont {R.~S.}\ \bibnamefont
  {Deacon}}, \bibinfo {author} {\bibfnamefont {S.}~\bibnamefont {Hartinger}},
  \bibinfo {author} {\bibfnamefont {O.}~\bibnamefont {Herrmann}}, \bibinfo
  {author} {\bibfnamefont {T.~M.}\ \bibnamefont {Klapwijk}}, \bibinfo {author}
  {\bibfnamefont {L.}~\bibnamefont {Maier}}, \bibinfo {author} {\bibfnamefont
  {C.}~\bibnamefont {Ames}}, \bibinfo {author} {\bibfnamefont {C.}~\bibnamefont
  {Brune}}, \bibinfo {author} {\bibfnamefont {C.}~\bibnamefont {Gould}},
  \bibinfo {author} {\bibfnamefont {A.}~\bibnamefont {Oiwa}}, \bibinfo {author}
  {\bibfnamefont {K.}~\bibnamefont {Ishibashi}}, \bibinfo {author}
  {\bibfnamefont {S.}~\bibnamefont {Tarucha}}, \bibinfo {author} {\bibfnamefont
  {H.}~\bibnamefont {Buhmann}}, \ and\ \bibinfo {author} {\bibfnamefont
  {L.~W.}\ \bibnamefont {Molenkamp}},\ }\href@noop {} {\bibfield  {journal}
  {\bibinfo  {journal} {Nature Communications}\ }\textbf {\bibinfo {volume}
  {7}},\ \bibinfo {pages} {10303} (\bibinfo {year} {2016})}\BibitemShut
  {NoStop}%
\bibitem [{\citenamefont {Albrecht}\ \emph {et~al.}(2016)\citenamefont
  {Albrecht}, \citenamefont {Higginbotham}, \citenamefont {Madsen},
  \citenamefont {Kuemmeth}, \citenamefont {Jespersen}, \citenamefont {Nygard},
  \citenamefont {Krogstrup},\ and\ \citenamefont {Marcus}}]{albrechtnat2016}%
  \BibitemOpen
  \bibfield  {author} {\bibinfo {author} {\bibfnamefont {S.~M.}\ \bibnamefont
  {Albrecht}}, \bibinfo {author} {\bibfnamefont {A.~P.}\ \bibnamefont
  {Higginbotham}}, \bibinfo {author} {\bibfnamefont {M.}~\bibnamefont
  {Madsen}}, \bibinfo {author} {\bibfnamefont {F.}~\bibnamefont {Kuemmeth}},
  \bibinfo {author} {\bibfnamefont {T.~S.}\ \bibnamefont {Jespersen}}, \bibinfo
  {author} {\bibfnamefont {J.}~\bibnamefont {Nygard}}, \bibinfo {author}
  {\bibfnamefont {P.}~\bibnamefont {Krogstrup}}, \ and\ \bibinfo {author}
  {\bibfnamefont {C.~M.}\ \bibnamefont {Marcus}},\ }\href@noop {} {\bibfield
  {journal} {\bibinfo  {journal} {Nature}\ }\textbf {\bibinfo {volume} {531}},\
  \bibinfo {pages} {206} (\bibinfo {year} {2016})}\BibitemShut {NoStop}%
\bibitem [{\citenamefont {Deacon}\ \emph {et~al.}(2016)\citenamefont {Deacon},
  \citenamefont {Wiedenmann}, \citenamefont {Bocquillon}, \citenamefont
  {Dominguez}, \citenamefont {Klapwijk}, \citenamefont {Leubner}, \citenamefont
  {Brune}, \citenamefont {Hankiewicz}, \citenamefont {Tarucha}, \citenamefont
  {Ishibashi}, \citenamefont {Buhmann},\ and\ \citenamefont
  {Molenkamp}}]{deaconarxiv2016}%
  \BibitemOpen
  \bibfield  {author} {\bibinfo {author} {\bibfnamefont {R.~S.}\ \bibnamefont
  {Deacon}}, \bibinfo {author} {\bibfnamefont {J.}~\bibnamefont {Wiedenmann}},
  \bibinfo {author} {\bibfnamefont {E.}~\bibnamefont {Bocquillon}}, \bibinfo
  {author} {\bibfnamefont {F.}~\bibnamefont {Dominguez}}, \bibinfo {author}
  {\bibfnamefont {T.~M.}\ \bibnamefont {Klapwijk}}, \bibinfo {author}
  {\bibfnamefont {P.}~\bibnamefont {Leubner}}, \bibinfo {author} {\bibfnamefont
  {C.}~\bibnamefont {Brune}}, \bibinfo {author} {\bibfnamefont {E.~M.}\
  \bibnamefont {Hankiewicz}}, \bibinfo {author} {\bibfnamefont
  {S.}~\bibnamefont {Tarucha}}, \bibinfo {author} {\bibfnamefont
  {K.}~\bibnamefont {Ishibashi}}, \bibinfo {author} {\bibfnamefont
  {H.}~\bibnamefont {Buhmann}}, \ and\ \bibinfo {author} {\bibfnamefont
  {L.~W.}\ \bibnamefont {Molenkamp}},\ }\href@noop {} {\bibfield  {journal}
  {\bibinfo  {journal} {arXiv}\ ,\ \bibinfo {pages} {1603.09611}} (\bibinfo
  {year} {2016})}\BibitemShut {NoStop}%
\bibitem [{\citenamefont {Shabani}\ \emph
  {et~al.}(2014{\natexlab{a}})\citenamefont {Shabani}, \citenamefont
  {Das~Sarma},\ and\ \citenamefont {Palmstr\o{}m}}]{shabaniprb2014}%
  \BibitemOpen
  \bibfield  {author} {\bibinfo {author} {\bibfnamefont {J.}~\bibnamefont
  {Shabani}}, \bibinfo {author} {\bibfnamefont {S.}~\bibnamefont {Das~Sarma}},
  \ and\ \bibinfo {author} {\bibfnamefont {C.~J.}\ \bibnamefont
  {Palmstr\o{}m}},\ }\href {\doibase 10.1103/PhysRevB.90.161303} {\bibfield
  {journal} {\bibinfo  {journal} {Phys. Rev. B}\ }\textbf {\bibinfo {volume}
  {90}},\ \bibinfo {pages} {161303} (\bibinfo {year}
  {2014}{\natexlab{a}})}\BibitemShut {NoStop}%
\bibitem [{\citenamefont {Shabani}\ \emph
  {et~al.}(2014{\natexlab{b}})\citenamefont {Shabani}, \citenamefont
  {McFadden}, \citenamefont {Shojaei},\ and\ \citenamefont
  {Palmstr\o{}m}}]{shabaniapl2014}%
  \BibitemOpen
  \bibfield  {author} {\bibinfo {author} {\bibfnamefont {J.}~\bibnamefont
  {Shabani}}, \bibinfo {author} {\bibfnamefont {A.~P.}\ \bibnamefont
  {McFadden}}, \bibinfo {author} {\bibfnamefont {B.}~\bibnamefont {Shojaei}}, \
  and\ \bibinfo {author} {\bibfnamefont {C.~J.}\ \bibnamefont {Palmstr\o{}m}},\
  }\href@noop {} {\bibfield  {journal} {\bibinfo  {journal} {Applied Physics
  Letters}\ }\textbf {\bibinfo {volume} {105}},\ \bibinfo {eid} {262105}
  (\bibinfo {year} {2014}{\natexlab{b}})}\BibitemShut {NoStop}%
\bibitem [{\citenamefont {Blonder}\ \emph {et~al.}(1982)\citenamefont
  {Blonder}, \citenamefont {Tinkham},\ and\ \citenamefont {Klapwijk}}]{BTK}%
  \BibitemOpen
  \bibfield  {author} {\bibinfo {author} {\bibfnamefont {G.~E.}\ \bibnamefont
  {Blonder}}, \bibinfo {author} {\bibfnamefont {M.}~\bibnamefont {Tinkham}}, \
  and\ \bibinfo {author} {\bibfnamefont {T.~M.}\ \bibnamefont {Klapwijk}},\
  }\href {\doibase 10.1103/PhysRevB.25.4515} {\bibfield  {journal} {\bibinfo
  {journal} {Phys. Rev. B}\ }\textbf {\bibinfo {volume} {25}},\ \bibinfo
  {pages} {4515} (\bibinfo {year} {1982})}\BibitemShut {NoStop}%
\bibitem [{\citenamefont {Irie}\ \emph {et~al.}(2016)\citenamefont {Irie},
  \citenamefont {Todt}, \citenamefont {Kumada}, \citenamefont {Harada},
  \citenamefont {Sugiyama}, \citenamefont {Akazaki},\ and\ \citenamefont
  {Muraki}}]{irieprb2016}%
  \BibitemOpen
  \bibfield  {author} {\bibinfo {author} {\bibfnamefont {H.}~\bibnamefont
  {Irie}}, \bibinfo {author} {\bibfnamefont {C.}~\bibnamefont {Todt}}, \bibinfo
  {author} {\bibfnamefont {N.}~\bibnamefont {Kumada}}, \bibinfo {author}
  {\bibfnamefont {Y.}~\bibnamefont {Harada}}, \bibinfo {author} {\bibfnamefont
  {H.}~\bibnamefont {Sugiyama}}, \bibinfo {author} {\bibfnamefont
  {T.}~\bibnamefont {Akazaki}}, \ and\ \bibinfo {author} {\bibfnamefont
  {K.}~\bibnamefont {Muraki}},\ }\href {\doibase 10.1103/PhysRevB.94.155305}
  {\bibfield  {journal} {\bibinfo  {journal} {Phys. Rev. B}\ }\textbf {\bibinfo
  {volume} {94}},\ \bibinfo {pages} {155305} (\bibinfo {year}
  {2016})}\BibitemShut {NoStop}%
\bibitem [{\citenamefont {Tang}\ \emph {et~al.}(1996)\citenamefont {Tang},
  \citenamefont {Wang},\ and\ \citenamefont {Zhang}}]{tangzpb1996}%
  \BibitemOpen
  \bibfield  {author} {\bibinfo {author} {\bibfnamefont {H.~X.}\ \bibnamefont
  {Tang}}, \bibinfo {author} {\bibfnamefont {Z.~D.}\ \bibnamefont {Wang}}, \
  and\ \bibinfo {author} {\bibfnamefont {Y.}~\bibnamefont {Zhang}},\
  }\href@noop {} {\bibfield  {journal} {\bibinfo  {journal} {Z. Phys. B}\
  }\textbf {\bibinfo {volume} {101}},\ \bibinfo {pages} {359} (\bibinfo {year}
  {1996})}\BibitemShut {NoStop}%
\bibitem [{\citenamefont {Burset}\ \emph {et~al.}(2014)\citenamefont {Burset},
  \citenamefont {Keidel}, \citenamefont {Tanaka}, \citenamefont {Nagaosa},\
  and\ \citenamefont {Trauzettel}}]{bursetprb2014}%
  \BibitemOpen
  \bibfield  {author} {\bibinfo {author} {\bibfnamefont {P.}~\bibnamefont
  {Burset}}, \bibinfo {author} {\bibfnamefont {F.}~\bibnamefont {Keidel}},
  \bibinfo {author} {\bibfnamefont {Y.}~\bibnamefont {Tanaka}}, \bibinfo
  {author} {\bibfnamefont {N.}~\bibnamefont {Nagaosa}}, \ and\ \bibinfo
  {author} {\bibfnamefont {B.}~\bibnamefont {Trauzettel}},\ }\href {\doibase
  10.1103/PhysRevB.90.085438} {\bibfield  {journal} {\bibinfo  {journal} {Phys.
  Rev. B}\ }\textbf {\bibinfo {volume} {90}},\ \bibinfo {pages} {085438}
  (\bibinfo {year} {2014})}\BibitemShut {NoStop}%
\bibitem [{\citenamefont {Burset}\ \emph {et~al.}(2015)\citenamefont {Burset},
  \citenamefont {Lu}, \citenamefont {Tkachov}, \citenamefont {Tanaka},
  \citenamefont {Hankiewicz},\ and\ \citenamefont
  {Trauzettel}}]{bursetprb2015}%
  \BibitemOpen
  \bibfield  {author} {\bibinfo {author} {\bibfnamefont {P.}~\bibnamefont
  {Burset}}, \bibinfo {author} {\bibfnamefont {B.}~\bibnamefont {Lu}}, \bibinfo
  {author} {\bibfnamefont {G.}~\bibnamefont {Tkachov}}, \bibinfo {author}
  {\bibfnamefont {Y.}~\bibnamefont {Tanaka}}, \bibinfo {author} {\bibfnamefont
  {E.~M.}\ \bibnamefont {Hankiewicz}}, \ and\ \bibinfo {author} {\bibfnamefont
  {B.}~\bibnamefont {Trauzettel}},\ }\href {\doibase
  10.1103/PhysRevB.92.205424} {\bibfield  {journal} {\bibinfo  {journal} {Phys.
  Rev. B}\ }\textbf {\bibinfo {volume} {92}},\ \bibinfo {pages} {205424}
  (\bibinfo {year} {2015})}\BibitemShut {NoStop}%
\bibitem [{\citenamefont {Hoppe}\ \emph {et~al.}(2000)\citenamefont {Hoppe},
  \citenamefont {Z\"ulicke},\ and\ \citenamefont {Sch\"on}}]{hoppeprl2000}%
  \BibitemOpen
  \bibfield  {author} {\bibinfo {author} {\bibfnamefont {H.}~\bibnamefont
  {Hoppe}}, \bibinfo {author} {\bibfnamefont {U.}~\bibnamefont {Z\"ulicke}}, \
  and\ \bibinfo {author} {\bibfnamefont {G.}~\bibnamefont {Sch\"on}},\ }\href
  {\doibase 10.1103/PhysRevLett.84.1804} {\bibfield  {journal} {\bibinfo
  {journal} {Phys. Rev. Lett.}\ }\textbf {\bibinfo {volume} {84}},\ \bibinfo
  {pages} {1804} (\bibinfo {year} {2000})}\BibitemShut {NoStop}%
\bibitem [{\citenamefont {Giazotto}\ \emph {et~al.}(2005)\citenamefont
  {Giazotto}, \citenamefont {Governale}, \citenamefont {Z\"ulicke},\ and\
  \citenamefont {Beltram}}]{giazottoprb2005}%
  \BibitemOpen
  \bibfield  {author} {\bibinfo {author} {\bibfnamefont {F.}~\bibnamefont
  {Giazotto}}, \bibinfo {author} {\bibfnamefont {M.}~\bibnamefont {Governale}},
  \bibinfo {author} {\bibfnamefont {U.}~\bibnamefont {Z\"ulicke}}, \ and\
  \bibinfo {author} {\bibfnamefont {F.}~\bibnamefont {Beltram}},\ }\href
  {\doibase 10.1103/PhysRevB.72.054518} {\bibfield  {journal} {\bibinfo
  {journal} {Phys. Rev. B}\ }\textbf {\bibinfo {volume} {72}},\ \bibinfo
  {pages} {054518} (\bibinfo {year} {2005})}\BibitemShut {NoStop}%
\bibitem [{\citenamefont {Khaymovich}\ \emph {et~al.}(2010)\citenamefont
  {Khaymovich}, \citenamefont {Chtchelkatchev}, \citenamefont {Shereshevskii},\
  and\ \citenamefont {Mel'nikov}}]{khaymovichel2010}%
  \BibitemOpen
  \bibfield  {author} {\bibinfo {author} {\bibfnamefont {I.~M.}\ \bibnamefont
  {Khaymovich}}, \bibinfo {author} {\bibfnamefont {N.~M.}\ \bibnamefont
  {Chtchelkatchev}}, \bibinfo {author} {\bibfnamefont {I.~A.}\ \bibnamefont
  {Shereshevskii}}, \ and\ \bibinfo {author} {\bibfnamefont {A.~S.}\
  \bibnamefont {Mel'nikov}},\ }\href
  {http://stacks.iop.org/0295-5075/91/i=1/a=17005} {\bibfield  {journal}
  {\bibinfo  {journal} {EPL (Europhysics Letters)}\ }\textbf {\bibinfo {volume}
  {91}},\ \bibinfo {pages} {17005} (\bibinfo {year} {2010})}\BibitemShut
  {NoStop}%
\bibitem [{\citenamefont {Snelder}\ \emph {et~al.}(2015)\citenamefont
  {Snelder}, \citenamefont {Stehno}, \citenamefont {Golubov}, \citenamefont
  {Molenaar}, \citenamefont {Scholten}, \citenamefont {Wu}, \citenamefont
  {Huang}, \citenamefont {van~der Wiel}, \citenamefont {Golden},\ and\
  \citenamefont {Brinkman}}]{snelderarxiv2015}%
  \BibitemOpen
  \bibfield  {author} {\bibinfo {author} {\bibfnamefont {M.}~\bibnamefont
  {Snelder}}, \bibinfo {author} {\bibfnamefont {M.~P.}\ \bibnamefont {Stehno}},
  \bibinfo {author} {\bibfnamefont {A.~A.}\ \bibnamefont {Golubov}}, \bibinfo
  {author} {\bibfnamefont {C.~G.}\ \bibnamefont {Molenaar}}, \bibinfo {author}
  {\bibfnamefont {T.}~\bibnamefont {Scholten}}, \bibinfo {author}
  {\bibfnamefont {D.}~\bibnamefont {Wu}}, \bibinfo {author} {\bibfnamefont
  {Y.~K.}\ \bibnamefont {Huang}}, \bibinfo {author} {\bibfnamefont {W.~G.}\
  \bibnamefont {van~der Wiel}}, \bibinfo {author} {\bibfnamefont {M.~S.}\
  \bibnamefont {Golden}}, \ and\ \bibinfo {author} {\bibfnamefont
  {A.}~\bibnamefont {Brinkman}},\ }\href@noop {} {\bibfield  {journal}
  {\bibinfo  {journal} {arXiv}\ }\textbf {\bibinfo {volume} {1506}},\ \bibinfo
  {pages} {05923} (\bibinfo {year} {2015})}\BibitemShut {NoStop}%
\bibitem [{\citenamefont {Woods}\ \emph {et~al.}(2004)\citenamefont {Woods},
  \citenamefont {Soulen}, \citenamefont {Mazin}, \citenamefont {Nadgorny},
  \citenamefont {Osofsky}, \citenamefont {Sanders}, \citenamefont {Srikanth},
  \citenamefont {Egelhoff},\ and\ \citenamefont {Datla}}]{woodsprb2004}%
  \BibitemOpen
  \bibfield  {author} {\bibinfo {author} {\bibfnamefont {G.~T.}\ \bibnamefont
  {Woods}}, \bibinfo {author} {\bibfnamefont {R.~J.}\ \bibnamefont {Soulen}},
  \bibinfo {author} {\bibfnamefont {I.}~\bibnamefont {Mazin}}, \bibinfo
  {author} {\bibfnamefont {B.}~\bibnamefont {Nadgorny}}, \bibinfo {author}
  {\bibfnamefont {M.~S.}\ \bibnamefont {Osofsky}}, \bibinfo {author}
  {\bibfnamefont {J.}~\bibnamefont {Sanders}}, \bibinfo {author} {\bibfnamefont
  {H.}~\bibnamefont {Srikanth}}, \bibinfo {author} {\bibfnamefont {W.~F.}\
  \bibnamefont {Egelhoff}}, \ and\ \bibinfo {author} {\bibfnamefont
  {R.}~\bibnamefont {Datla}},\ }\href {\doibase 10.1103/PhysRevB.70.054416}
  {\bibfield  {journal} {\bibinfo  {journal} {Phys. Rev. B}\ }\textbf {\bibinfo
  {volume} {70}},\ \bibinfo {pages} {054416} (\bibinfo {year}
  {2004})}\BibitemShut {NoStop}%
\bibitem [{\citenamefont {DeGrave}\ \emph {et~al.}(2011)\citenamefont
  {DeGrave}, \citenamefont {Schmitt}, \citenamefont {Selinsky}, \citenamefont
  {Higgins}, \citenamefont {Keavney},\ and\ \citenamefont
  {Jin}}]{degravenl2011}%
  \BibitemOpen
  \bibfield  {author} {\bibinfo {author} {\bibfnamefont {J.~P.}\ \bibnamefont
  {DeGrave}}, \bibinfo {author} {\bibfnamefont {A.~L.}\ \bibnamefont
  {Schmitt}}, \bibinfo {author} {\bibfnamefont {R.~S.}\ \bibnamefont
  {Selinsky}}, \bibinfo {author} {\bibfnamefont {J.~M.}\ \bibnamefont
  {Higgins}}, \bibinfo {author} {\bibfnamefont {D.~J.}\ \bibnamefont
  {Keavney}}, \ and\ \bibinfo {author} {\bibfnamefont {S.}~\bibnamefont
  {Jin}},\ }\href {\doibase 10.1021/nl2026426} {\bibfield  {journal} {\bibinfo
  {journal} {Nano Letters}\ }\textbf {\bibinfo {volume} {11}},\ \bibinfo
  {pages} {4431} (\bibinfo {year} {2011})},\ \bibinfo {note} {pMID: 21923114},\
  \Eprint {http://arxiv.org/abs/http://dx.doi.org/10.1021/nl2026426}
  {http://dx.doi.org/10.1021/nl2026426} \BibitemShut {NoStop}%
\bibitem [{\citenamefont {Lippmann}\ and\ \citenamefont
  {Kuhrt}(1958)}]{lippmannzna1958}%
  \BibitemOpen
  \bibfield  {author} {\bibinfo {author} {\bibfnamefont {H.~J.}\ \bibnamefont
  {Lippmann}}\ and\ \bibinfo {author} {\bibfnamefont {R.}~\bibnamefont
  {Kuhrt}},\ }\href@noop {} {\bibfield  {journal} {\bibinfo  {journal} {Z.
  Naturforsch. A}\ }\textbf {\bibinfo {volume} {13}},\ \bibinfo {pages} {462}
  (\bibinfo {year} {1958})}\BibitemShut {NoStop}%
\bibitem [{\citenamefont {Jensen}\ and\ \citenamefont
  {Smith}(1972)}]{jensenjpc1972}%
  \BibitemOpen
  \bibfield  {author} {\bibinfo {author} {\bibfnamefont {H.~H.}\ \bibnamefont
  {Jensen}}\ and\ \bibinfo {author} {\bibfnamefont {H.}~\bibnamefont {Smith}},\
  }\href {http://stacks.iop.org/0022-3719/5/i=20/a=006} {\bibfield  {journal}
  {\bibinfo  {journal} {Journal of Physics C: Solid State Physics}\ }\textbf
  {\bibinfo {volume} {5}},\ \bibinfo {pages} {2867} (\bibinfo {year}
  {1972})}\BibitemShut {NoStop}%
\bibitem [{\citenamefont {Abanin}\ and\ \citenamefont
  {Levitov}(2008)}]{abanin2008}%
  \BibitemOpen
  \bibfield  {author} {\bibinfo {author} {\bibfnamefont {D.~A.}\ \bibnamefont
  {Abanin}}\ and\ \bibinfo {author} {\bibfnamefont {L.~S.}\ \bibnamefont
  {Levitov}},\ }\href {\doibase 10.1103/PhysRevB.78.035416} {\bibfield
  {journal} {\bibinfo  {journal} {Phys. Rev. B}\ }\textbf {\bibinfo {volume}
  {78}},\ \bibinfo {pages} {035416} (\bibinfo {year} {2008})}\BibitemShut
  {NoStop}%
\end{thebibliography}

%

\end{document}